\documentclass[11pt]{article}%
\usepackage{amsfonts}
\usepackage{amssymb}
\usepackage{caption,subcaption}
\usepackage{graphicx}
\usepackage{setspace}
\usepackage[round]{natbib}
\usepackage{amsmath}%
\usepackage{amsthm}%
\usepackage{hyperref}
\usepackage{url}
\usepackage{mathpazo}
\usepackage{paralist}
\usepackage{mathtools}
\usepackage{geometry}
\usepackage{booktabs}
\usepackage{microtype}

\newcommand{\R}{\mathcal{R}}

\newcommand{\bbeta}{\boldsymbol\beta}

\newcommand{\vnorm}[1]{\left|\left|#1\right|\right|}

\newcommand{\bx}{{\bf x}}

\title{Multiscale spatial density smoothing: an application to large-scale radiological survey and anomaly detection}
\author{Wesley Tansey\thanks{Department of Computer Science, University of Texas
    at Austin.}
  \and
  Alex Athey\thanks{Applied Research Laboratories, University of Texas at Austin.}
  \and
  Alex Reinhart\thanks{Department of Statistics, Carnegie Mellon University,
   \href{mailto:areinhar@stat.cmu.edu}{areinhar@stat.cmu.edu} (corresponding
   author).}
 \and
 James G.~Scott\thanks{Department of Information, Risk, and Operations
   Management; Department of Statistics and Data Sciences, University of Texas
   at Austin.}}
\date{This version: \today}

\begin{document}

\maketitle
\begin{abstract}
\noindent We consider the problem of estimating a spatially varying density function, motivated by problems that arise in large-scale radiological survey and anomaly detection.  In this context, the density functions to be estimated are the background gamma-ray energy spectra at sites spread across a large geographical area, such as nuclear production and waste-storage sites, military bases, medical facilities, university campuses, or the downtown of a city.  Several challenges combine to make this a difficult problem.  First, the spectral density at any given spatial location may have both smooth and non-smooth features.  Second, the spatial correlation in these density functions is neither stationary nor locally isotropic.  Finally, at some spatial locations, there is very little data.  We present a method called multiscale spatial density smoothing that successfully addresses these challenges.  The method is based on recursive dyadic partition of the sample space, and therefore shares much in common with other multiscale methods, such as wavelets and P\'olya-tree priors.  We describe an efficient algorithm for finding a maximum a posteriori (MAP) estimate that leverages recent advances in convex optimization for non-smooth functions.

We apply multiscale spatial density smoothing to real data collected on the background gamma-ray spectra at locations across a large university campus.  The method exhibits state-of-the-art performance for spatial smoothing in density estimation, and it leads to substantial improvements in power when used in conjunction with existing methods for detecting the kinds of radiological anomalies that may have important consequences for public health and safety.

\bigskip


\end{abstract}

\newpage

\section{Detecting radiation anomalies}
\label{sec:introduction}

\subsection{Introduction}

Lost or stolen radioactive sources present a challenging security problem. Widely used for industrial radiography, sterilization, and medical imaging, these sources are often poorly secured \citep{Gaffigan:2012vg} and sometimes stolen \citep{Korshukin:2006}. To prevent dangerous accidents and to detect radiological dispersal devices (dirty bombs) before they can be used, security agencies are interested in continuously monitoring wide areas for radiation sources. A simple method is to monitor overall radiation levels.  But these vary naturally in space, as different soil, stone, and building materials can contain widely different amounts of naturally occurring radioactive materials (NORM).  Moreover, different detectors exhibit very different overall sensitivities to radiation.  It is therefore more effective to monitor the energy spectrum of the detected radiation instead, as different radioactive materials emit gamma radiation at distinct energies.

But to find a spectral anomaly, one must first know what the normal background spectrum looks like at all spatial locations.  The background spectrum is the probability distribution for the energy of a gamma ray emitted by natural sources of radiation at a given site.  Therefore, a key statistical problem in radiological anomaly detection is the need to accurately estimate a spatially varying probability distribution for a set of random variables $X^{(s)}$, where $s$ indexes the site.

We present a new approach to this problem, called multiscale spatial density smoothing, that exploits a recursive dyadic partition of the sample space of $X^{(s)}$.  This re-casts the problem of estimating a density as the problem of estimating a set of binomial probabilities associated with the nodes in a binary tree.  We incorporate a prior on each binomial probability that encourages spatial smoothing, and we compute a fast MAP estimate under this model via a recently developed algorithm for graph-based total-variation denoising.  Our studies show that the method is fast, highly accurate for densities with difficult multiscale structure, and scalable to data sets too large to be stored on a single machine.  Our main application also shows that the method leads to cutting-edge performance in detecting radiological anomalies, using data on background gamma radiation collected over a period of six weeks on a large university campus.

\subsection{Background on radiological survey}

\paragraph{Our data.}  The data analysis described in this paper is part of a broader effort to assist first responders, who lack the capability to detect and monitor excessive levels of radiation, particularly over large geographic areas. Current methods are reactive, searching for radioactive sources if they are known to be lost, rather than proactive, monitoring continuously to detect unexpected sources. In collaboration with local public agencies, we are developing a system for monitoring radiation in complex urban areas using mobile detectors employed on existing infrastructure, such as garbage trucks, city buses, taxis, or mail carriers.  The goal is to develop an end-to-end system for monitoring and anomaly detection that does not require operator interface, a large operator equipment burden, or specialized training.


The data analyzed in this paper was collected as part of the initial pilot study with this system during July and August 2012 at the University of Texas J.J.~Pickle Research Campus (PRC), a satellite campus that houses various research facilities of the University of Texas at Austin.  Several features at the PRC cause its background spectrum to vary spatially: several brick buildings have higher background rates, and there is a radiological storage site on the northwest corner of the campus. By collecting data repeatedly over the course of six weeks, we are able to accurately map the background radiation spectrum and observe its change over time, detecting any unusual spectral changes.

To collect our data, we used a $2\times 2$ inch cesium iodide gamma spectrometer. The spectrometer continuously recorded gamma rays, binning them into 4,096 discrete energy bins.\footnote{Because energy bins are discrete, we are not estimating a density in the strict sense, merely a very fine histogram.  Nonetheless, to avoid needless complexity, we refer to ``densities'' and ``density smoothing'' .}  A laptop recorded the gamma spectrum, GPS location, and timestamp every two seconds. The detector typically recorded about 30--50 gamma rays per second. The detector was driven around campus on a golf cart once or twice daily, recording a total of 18 hours of observations taken over 41 drives on 30 different days.  See Appendix \ref{app:remarks} for further details on the data and pre-processing.

\paragraph{Existing work on radiological survey.} In addition to their law-enforcement applications, radiological surveys are important in many other security and environmental scenarios.  In medicine, they can be used to find a lost source, e.g.~a medical radio-isotope that has been stolen or gone missing from a hospital.  In border security, they can be used as part of a protocol to scan for illicit radiological material on trucks at a border crossing or container ships at a port.  In disaster response, they can be used to help assess the radiological impact of nuclear accidents such as Chernobyl or Fukushima \citep{Craig:2012}.

Previous source search methods have focused on detecting radioactive sources with a mobile detector when a previous radiological survey does not exist \citep{Pfund:2007,Pfund:2010hm}. Without previous knowledge of the background radiation spectrum, these systems must distinguish changes caused by man-made sources from natural variation in the background, reducing their sensitivity. Automated portal monitoring systems, for use at border crossings or ports, have been developed \citep{Runkle:2009ev}, and portal monitors are now widely deployed. However, though some cities have performed helicopter-based radiation surveys for disaster preparedness \citep{Wasiolek:2007vn}, these are impractically expensive for long-term radiation surveillance.

Previous work on long-term mapping and anomaly detection includes \citet{Aucott:2013bm}, which attempted to map the background and correlate spectral features with features detected by cameras and LIDAR, such as nearby buildings and construction materials; \citet{reinhart:etal:2014}, which used spectral comparison ratios to detect spectral changes in large $250\times 250$ meter spatial cells; and \citet{Reinhart:2015}, which used Kolmogorov--Smirnov tests to achieve higher power with much smaller spatial cells, giving a better ability to detect and localize sources. The latter two methods do not borrow information across space: each cell is treated as independent of the others, so each cell requires sufficient background observations to obtain a good background spectrum estimate. The purpose of this article is to show how a new methodology can address this shortcoming and improve overall detection performance.

\subsection{Statistical background}

\begin{figure}
\begin{center}
\includegraphics[width=\textwidth]{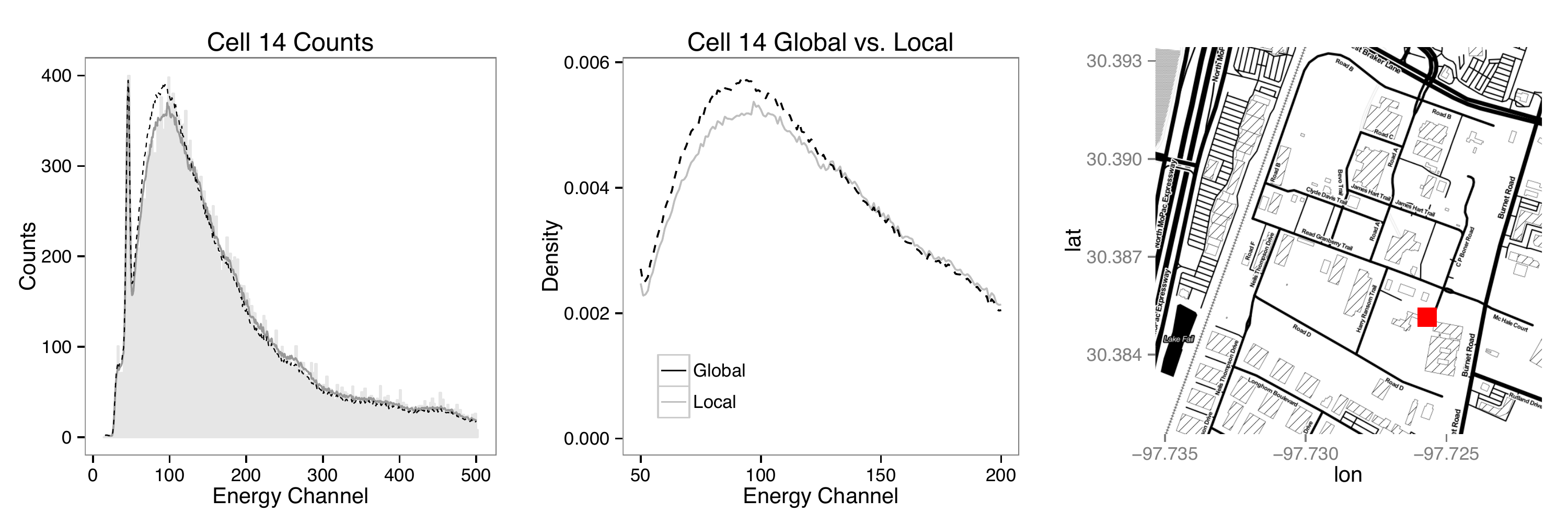} \\
\includegraphics[width=\textwidth]{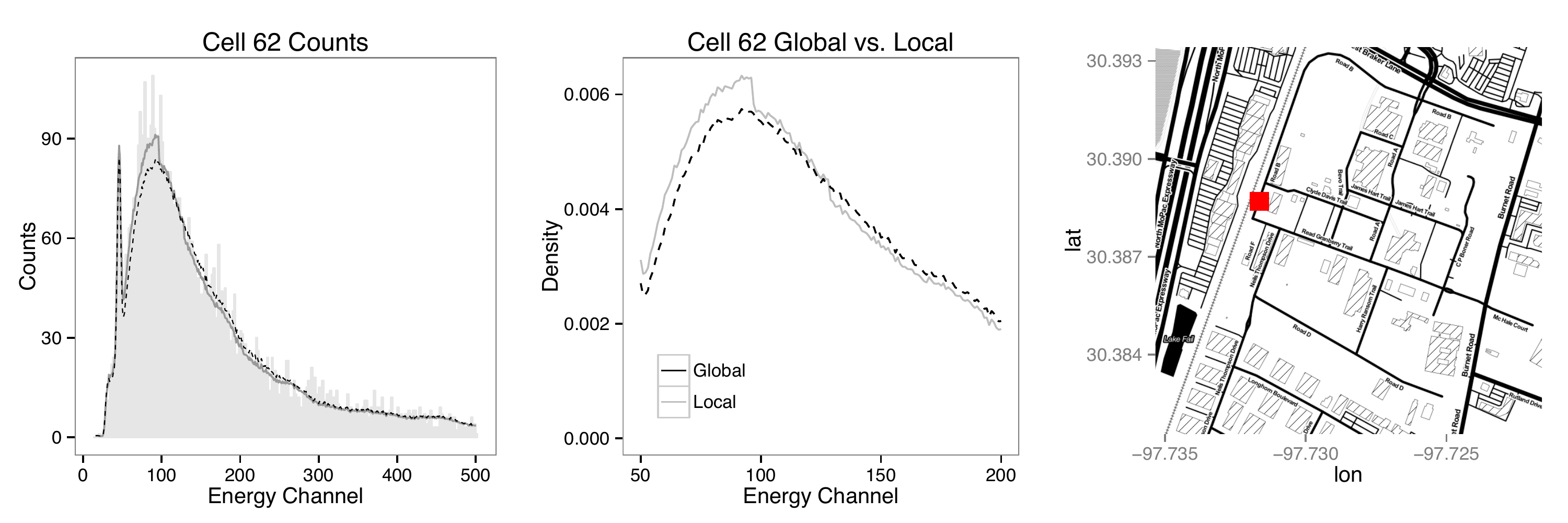}
\caption{\label{fig:prc_reconstruction_examples} Two example spatial sites (cell 14, top row; and cell 62, bottom row) for the PRC data that show the comparison between the global density estimate and the local density estimates using our procedure for multiscale spatial density smoothing.  The left two panels compare the training data in each cell (grey histogram); the local density estimates based on multiscale spatial smoothing (solid grey curve); and the spatially invariant global density estimate (dashed black curve).  (These densities are rescaled to show the expected count rate.)  The middle two panels show the difference between the global and local estimates near the broad peak in the spectra between channels 50 and 200.  The right two panels show the locations of each site on a larger map of campus.}
\end{center}
\end{figure}

Several factors make our problem difficult.  First, the background spectra we are trying to estimate have both smooth and spiky features. They are not well described by any simple class of density functions or by a single bandwidth parameter.  Outside a spatial context, this is often addressed by locally adaptive density estimation \citep[e.g.][]{sain:scott:1996}; it is a difficult problem in its own right, independent of the complication introduced by spatial variation.  Second, the spatial variation in background spectra is nonstationary and locally anisotropic.  One reason for this is that different building materials have different combinations of naturally radioactive isotopes, and there is wide variation in building materials across sites: some radioactive isotopes are common, while others are rare \citep{Ryan:2011wh}.  Finally, at some sites, there is very little data, because the background data-collection regime did not often visit these regions. This data sparsity means that accurate spatial interpolation of the background spectrum will often be necessary for the purposes of detecting anomalies in rarely visited sites.

Our proposed multiscale density-smoothing method successfully addresses these challenges.   Conceptually, it involves three steps:
\begin{enumerate}
\item Split the overall problem into sub-problems by recursively partitioning the sample space into a series of half-spaces, each described by a conditional probability.
\item Smooth the half-space probabilities by computing a fast MAP estimate under a prior that encourages spatial neighbors to be similar.
\item Merge the smoothed half-space probabilities to yield estimates $\hat{f}^{(s)}$ at each site.
\end{enumerate}

This ``split/smooth/merge'' structure reduces the density smoothing problem to a set of separate scalar smoothing problems, each of which can be solved by convex optimization.  It is therefore extremely fast and scalable to very large data sets, and can exploit two complementary strategies for parallelization. First, data for different spatial sites can be stored on different machines, with minimal communication between them.  Second, different scalar smoothing problems, corresponding to different half-space probabilities, can be solved on different machines and are easily combined to form a final answer.  We exploit both sources of parallelism in our analysis.  With these savings, analyzing enormous radiological data sets becomes computationally feasible.  

Figure \ref{fig:prc_reconstruction_examples} provides a snapshot of the method's performance on our data, as well as some intuition regarding the statistical challenges we have mentioned.  The figure shows the density reconstruction for two sites (labeled 14 and 62) for which a relatively large amount of data is available (because the map has been divided into a very coarse grid).  The raw gamma-ray counts for these two sites exhibit visibly different backgrounds near the broad peak in the spectrum that occurs between energy channels 50 and 200.  Our procedure faithfully reconstructs these differences without the excessive variability of the raw histograms, yet it still manages to capture other ``spiky'' features of the density without oversmoothing.

\paragraph{Connection with existing work.} Our method builds on two large, independent bodies of statistical literature: density estimation and spatial smoothing.  

There is a significant body of work on conditional density estimation, of which spatially varying density estimation is a subset.  Most frequentist work on this subject is based on kernel methods \citep[see, e.g.][and the references therein]{bash:hynd:2001}.  However, these techniques typically involve estimating a density that changes smoothly as a function of covariates.  They are not directly suited to the spatial context, and in any case, the spatial variation in our data is not smooth.  Moreover, traditional (non-adaptive) kernel methods do poorly at estimating densities with multiscale structure like those we face, which contain both spiky and smooth features, both in the spatial and energy domains.  No off-the-shelf kernel method seems appropriate for our problem.  There is also a substantial body of Bayesian work on conditional density estimation using the dependent Dirichlet process and its variations \citep{maceachern:2000}.  But these methods scale poorly, which makes them intractably slow for our problem.

The technique of representing a probability distribution via a recursive dyadic partition has been widely used---for example, in multiscale models for Poisson and multinomial estimation \citep{fryzlewicz:nason:2004,jansen:2006,willett:nowak:2007}, and in nonparametric Bayesian inference via P\'olya-tree priors \citep[e.g.][]{mauldin:sudderth:williams:1992,lavine:1992,lavine:1994}.   Conceptually, our paper is most closely related to the proposal by \citet{zhou:hanson:2011}, who couple a Polya-tree prior with a conditional autoregressive (CAR) prior to estimate a spatially varying survival model.  However, theirs is a fully Bayesian model and must be estimated by Markov-chain Monte Carlo.  This scales poorly to very large spatial lattices of the kind necessary to characterize fine-scale radiological features across an entire city.  Our method is not fully Bayesian, and it is fit by convex optimization, which scales much more favorably.  Moreover, on experiments that are small enough for the method of \citet{zhou:hanson:2011} to be fit, our method has comparable or better accuracy, but is orders of magnitude faster.  The price we pay for scalability is that, unlike a fully Bayes approach, our method does not provide error bars---only a point estimate.

Spatial smoothing is a large and mature field, and we do not attempt a comprehensive review.  Our approach is closely related to a technique called total-variation denoising from the image-processing literature \citep{rudin:osher:fatemi:1992}, which is also known in statistics as the fused lasso \citep{tibs:fusedlasso:2005}.  We draw heavily on recent work about computationally efficient estimation for this class of optimization problems, including \citet{tibs:taylor:2011}, \citet{ramdas:tibs:2014}, \citet{wahlberg:etal:2012}, \citet{tansey:etal:2014}, \citet{wang2014trend}, and \citet{tansey:scott:2015}.  Specifically, we use the algorithm from \citet{tansey:scott:2015} \citep[which is itself strongly motivated by the discussion in][]{wang2014trend} to solve a series of optimization problems that combine a binomial likelihood together with a total-variation penalty over the nodes of an undirected graph.  Our approach is also conceptually related to Bayesian inference for areal data models.  Specifically, it is similar to conditional auto-regressive (CAR) models \citep{besag74}, which also effect spatial smoothing by discouraging large pairwise differences across edges in a graph.  Nonparametric Bayesian approaches for spatial smoothing were investigated by \citet{gelfandkottas2005}, \citet{reich:fuentes:2007}, and \citet{rodriguez:etal:2010}, among many others, and we refer the interested readers to these papers, and their references, for more detail.

\subsection{Outline}

Section \ref{sec:preliminaries} describes the overall structure of our ``split/smooth/merge'' approach.  Section \ref{sec:spatial_smoothing} details our approach to spatial smoothing and our fitting algorithm.  Section \ref{sec:simulations} presents simulation evidence that our method is highly effective at estimating a spatially varying density function.  Section \ref{sec:anomaly_example} describes our main application to radiological anomaly detection.  Section \ref{sec:conclusions} concludes with some final remarks.

\section{Overview of approach}
\label{sec:preliminaries}

\subsection{Notation}

In this paper, space is assumed to be discrete and represented by an undirected graph $\mathcal{G} = (\mathcal{V}, \mathcal{E})$, where an edge $(r,s)$ means that sites $r$ and $s$ are neighbors.  We use $p = |\mathcal{V}|$ to denote the number of vertices and $d = |\mathcal{E}|$ the number of edges.  Let $f^{(s)}$ be the set of densities across sites, all with common sample space $B$.  At each site, we observe data $\bx^{(s)} = (x^{(s)}_1, \ldots, x^{(s)}_{N^{(s)}})$, a vector of observations $x^{(s)}_{i} \stackrel{i.i.d.}{\sim} f^{(s)}$ of length $N^{(s)}$.

A key device in our paper is the use of a recursive dyadic partition of the sample space $B$.  We recursively construct the level-$k$ partition of $B$, denoted $\Pi^{(k)}$, via a bijection with all length-$k$ binary sequences $\gamma \in \{0,1\}^k$, as follows.  Let the level-$1$ partition as $\Pi^{(1)} = \{B_0, B_1\}$ where $B_0 \cup B_1 = B$ and $B_0 \cap B_1 = \emptyset$.  Given the partition at level $k$, the partition at level $k+1$ is defined by specifying, for each $\gamma \in \{0,1\}^k$, the pair $(B_{\gamma 0}, B_{\gamma 1})$ satisfying $B_{\gamma 0} \cup B_{\gamma 1} = B_{\gamma}$ and $B_{\gamma 0} \cap B_{\gamma 1} = \emptyset$.  Here $\gamma0$ ($\gamma1$) is new binary sequence defined by appending a 0 (1) to the end of $\gamma$.

\begin{figure}[t]
\centering
\begin{subfigure}[t]{3in}
\caption{One density}
\includegraphics[width=3in]{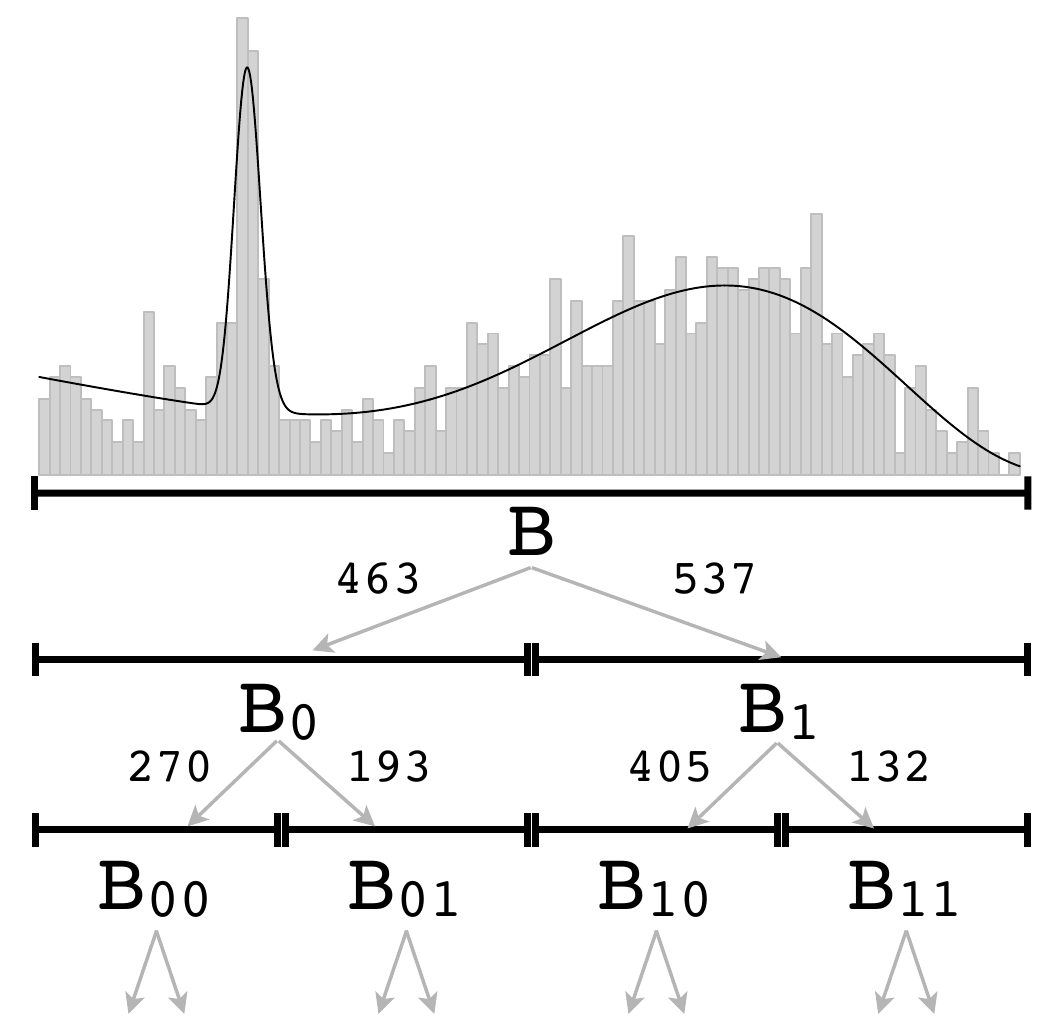}
\end{subfigure}
\begin{subfigure}[t]{3in}
\caption{Spatially varying density}
\includegraphics[width=3in]{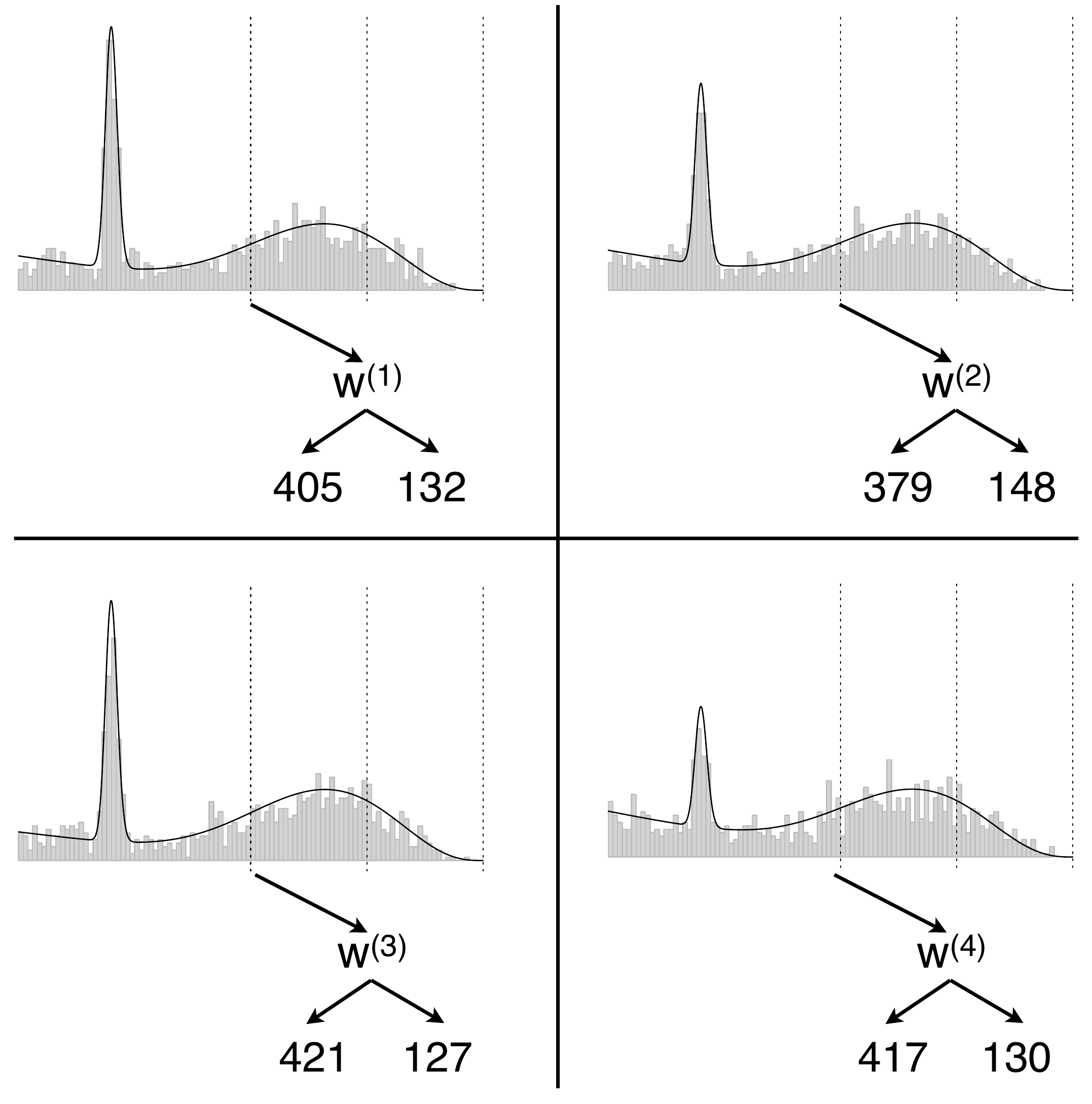}
\end{subfigure}
\caption{\label{fig:rdp_example} (Panel A) An example of a depth-2 recursive dyadic partition of an interval $B$, together with a sample $\bx$ of size $1000$ from a known density (shown as a thin black line).  The counts next to each arrow show the quantity $y(\bx, B_{\gamma})$.  For example, $y(\bx, B_{0}) = 270$, because 270 of the 463 points in $B_0$ fall in the left child node $B_{00}$.  (Panel B) A toy example of a spatial smoothing problem for the splitting node $B_1$ from Panel A, where the underlying density differs at each site in a $2 \times 2$ grid graph.  (The density from Panel A is at the top-left site.)  The superscripts on the $w$'s index spatial sites, not nodes in the tree.}
\end{figure}

We refer to $B_{\gamma}$ (or just $\gamma$) as a parent node, $B_{\gamma0}$ as the left child, and $B_{\gamma1}$ as the right child.   To give a concrete example, if $B$ is the unit interval (i.e.~the level-0 partition), the level-$1$ partition could be $\Pi^{(1)} = \{[0,0.5], (0.5, 1]\}$; the level-$2$ partition could be $\Pi^{(2)} = \{[0, 0.25], (0.25, 0.5], (0.5, 0.75], (0.75, 1]\}$;  and so on.   (See Panel A, Figure \ref{fig:rdp_example}.)

Together with a splitting rule for constructing the child nodes $B_{\gamma 0}$ and $B_{\gamma 1}$ from $B_{\gamma}$, this recursive process defines a depth-$K$ binary tree whose complete set of nodes is
$$
\mathcal{B}^{(K)} = \bigcup_{k=1}^K \Pi^{(k)} = \bigcup_{k=1}^K  \big( B_{\gamma} : \gamma \in \{0,1\}^k \big) \, ,
$$
indexed by all binary sequences $\gamma$ of length no more than $K$.  There are many possible splitting rules; we choose $B_{\gamma 0}$ and $B_{\gamma 1}$ to have equal length, assuming that $B$ is compact.

\subsection{Split, smooth, and merge}

This tree $\mathcal{B}^{(K)}$ enables a ``split/smooth/merge'' estimation strategy, as follows.  For now we drop the spatial index $s$, and consider the density $f$ at a single site (e.g.~Panel A of Figure \ref{fig:rdp_example}).  Suppose that $X \sim f$, and that $B_{\gamma}$ is some (nonterminal) node in $\mathcal{B}^{(K)}$.  We focus on the conditional probability that $X$ falls in the left-child node $B_{\gamma0}$, given that it is in $B_{\gamma}$:
$$
w_{\gamma} = P(X \in B_{\gamma0} \mid X \in B_{\gamma})
$$
We then express the probabilities at the terminal nodes of the tree as the product of the $w_{\gamma}$'s as one traverses the tree starting at the root.  These terminal-node probabilities form a very fine discrete approximation to $f$.

Now suppose that $\bx = (x_1, \ldots, x_n)$ is a vector of $n$ i.i.d.~samples from $f$, and let $y(\bx; b) = \#\{x_i \in b\}$ be the number of these samples that fall in the set $b \subset B$. If $B_{\gamma}$ is a nonterminal node, then clearly
\begin{equation}
\label{eqn:binomial_likelihood}
(y_{\gamma0} \mid y_{\gamma}) \sim \mbox{Binomial}(y_{\gamma}, w_{\gamma}) \, ,
\end{equation}
where $y_{\gamma} = y(\bx, B_{\gamma})$ for short.  In this way, we have reduced the density-estimation problem to a set of independent binomial problems, one for each node in $\mathcal{B}^{(K)}$.  Crucially, each $w_{\gamma}$ can be estimated independently.

This explains the ``split'' and ``merge'' aspects of our estimation strategy.  In the split step, we form separate binomial-estimation problems for each $w_{\gamma}$.  In the merge step, after having estimated the $w_{\gamma}$'s, we recursively compute the terminal-node probabilities from the conditional splitting probabilities.  The missing ingredient here is spatial smoothing.  To explain this, we re-introduce the spatial index $s$ and consider the collection of splitting probabilities $w_{\gamma}^{(s)}$ and counts $y_{\gamma}^{(s)}$ for all tree nodes $\gamma$ and sites $s \in \mathcal{V}$.  Panel B of Figure \ref{fig:rdp_example} shows a toy example for the node $\gamma = 1$, where the underlying density differs for each site in a $2 \times 2$ grid.  Concretely, at site 3, we have $y_{10} = 421$ and $y_{11} = 127$.

Our method applies spatial smoothing to the batch of splitting probabilities $W_{\gamma} = \big\{{w}^{(s)}_{\gamma} : s \in \mathcal{V} \big\}$ associated with a single splitting node $\gamma$, across all spatial sites, before the ``merge'' step.  We do this using the binomial likelihood defined by Equation (\ref{eqn:binomial_likelihood}) together with a penalty function that encourages spatial smoothness.  The rest of the paper will describe this step in detail.


\section{Spatial smoothing via graph-based denoising}
\label{sec:spatial_smoothing}

\subsection{Penalizing the total variation on a graph}

By construction, the batch of splitting probabilities for one tree node, $\big\{ {w}^{(s)}_{\gamma} : s \in \mathcal{V} \big\}$, can be estimated separately from the batch for another tree node.  Therefore we consider the smoothing problem for a single nonterminal node $B_{\gamma} \in \mathcal{B}$, and we drop the index $\gamma$.  Let $y^{(s)} \equiv y(\bx^{(s)}, B_{\gamma0})$ be the count in the left child at site $s$, and let $m^{(s)} \equiv y(\bx^{(s)}, B_{\gamma})$ be the count in the parent.    The left-child count follows a binomial distribution, given the parent count: $(y^{(s)} \mid m^{(s)}, w^{(s)}) \sim \mbox{Binomial}(m^{(s)}, w^{(s)})$. The goal is to estimate the conditional left-split probability $w^{(s)}$ at every site.

We take a penalized-likelihood approach, by penalizing the total variation (TV) of the log-odds across the graph $\mathcal{G}$.  This is also known as the graph fused lasso, or zeroth-order graph trend filtering \citep{wang2014trend}.  Let $\beta^{(s)} = \log\{w^{(s)}/(1-w^{(s)})\}$ denote the log-odds corresponding to the left-split probability at site $s$.  We estimate $\{\beta^{(s)} : s \in \mathcal{V}\}$ by solving the following optimization problem:
\begin{equation}
\label{eqn:logit_objective0}
\begin{aligned}
& \underset{\beta \in \R^p}{\text{minimize}}
& & 
\sum_{s \in \mathcal{V}} \left[ m^{(s)} \log \left\{ 1+ \exp (\beta^{(s)}) \right\} - y^{(s)} \beta^{(s)} \right ] + \lambda \sum_{(r,s) \in \mathcal{E}} |\beta^{(r)} - \beta^{(s)}| \, ,
\end{aligned}
\end{equation}
where $\lambda$ is a penalty parameter chosen adaptively (see Section \ref{sec:algorithm_details}).  This is also known as the graph-fused lasso (GFL) under a binomial likelihood.  To reiterate: this is the problem for one tree node.  We solve a separate such problem for every non-terminal node, in parallel.


The choice of an $\ell^1$ penalty function can be motivated on both computational and statistical grounds.  First, it leads to a convex optimization problem.  The computational advantages of convexity are an important consideration, since problem (\ref{eqn:logit_objective0}) must be solved hundreds or thousands of times (once for each tree split) on a large spatial lattice.  Second, the $\ell^1$ penalty leads to a nonlinear smoother, capable of adapting to varying degrees of smoothness in different spatial regions.  This contrasts with the linear smoothing provided by an $\ell^2$ penalty (or analogously, a Gaussian CAR prior).  This adaptivity has important consequences for estimation accuracy, both theoretically and practically.  For example, \citet{sadhanala:etal:2016} have established minimax rates for  estimating functions on regular grids under squared-error loss, extending the work of \citet{donoho:johnstone:1998} on 1D functions. They showed that, while total-variation denoising achieves the minimax rate, all linear smoothers (including $\ell^2$-based smoothers) fail to do so.  These results do not apply directly to our problem, which involves neither a regular grid nor squared-error loss.  But they are suggestive that the $\ell^1$ penalty may be preferred to the $\ell^2$ penalty for spatial smoothing, unless we believe that the underlying spatial field is homogeneously smooth.  (In our application, there are strong reasons to believe the opposite, due to line-of-sight occlusions that can scatter gamma rays from physically nearby sources.)

\begin{figure}
\centering
\vspace{-0.3in}
\begin{subfigure}{4.8in}
\caption{Truth is smooth\vspace{-0.1in}}
\includegraphics[width=4.9in]{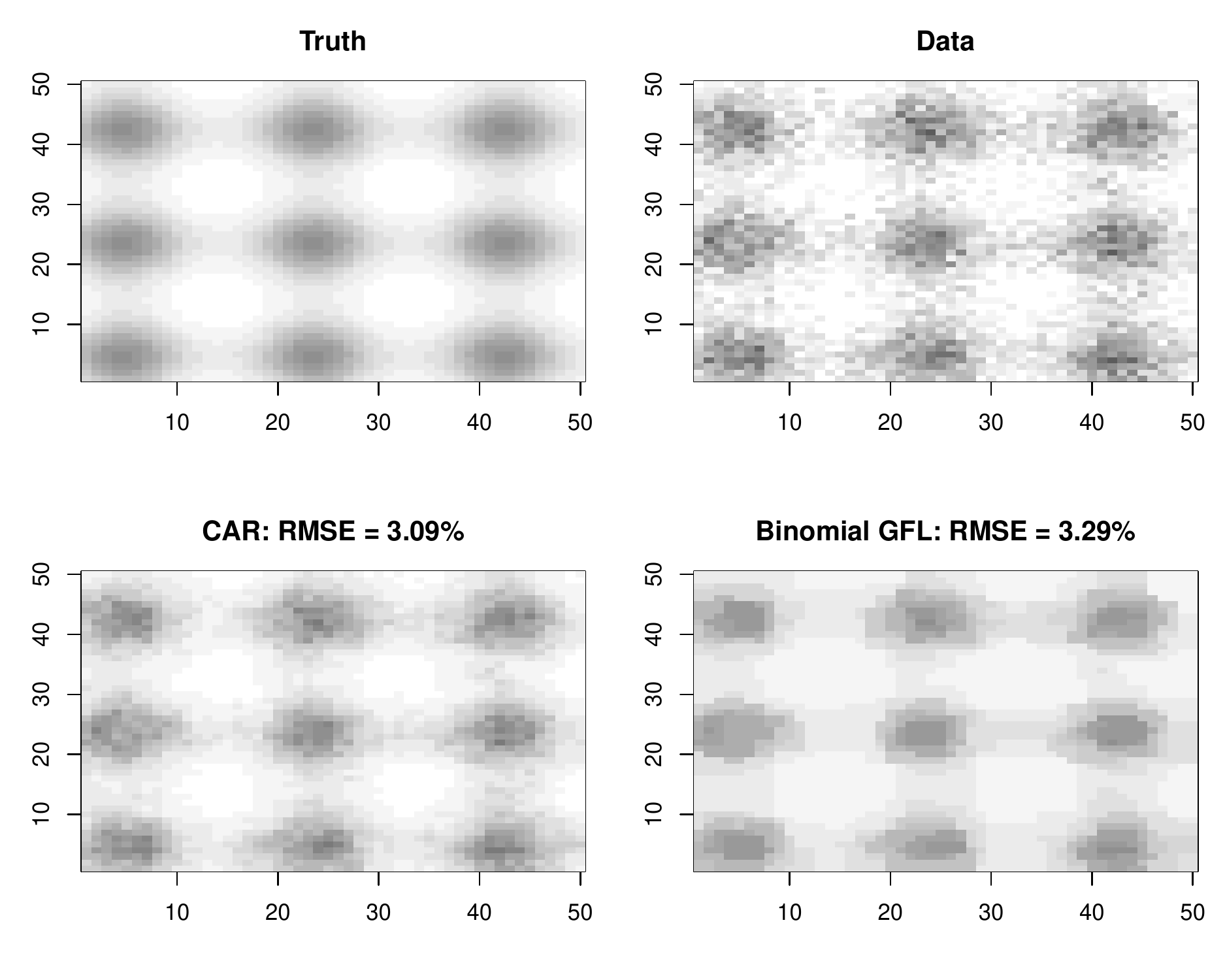}
\end{subfigure}
\begin{subfigure}{4.8in}
\caption{Truth is not smooth\vspace{-0.1in}}
\includegraphics[width=4.9in]{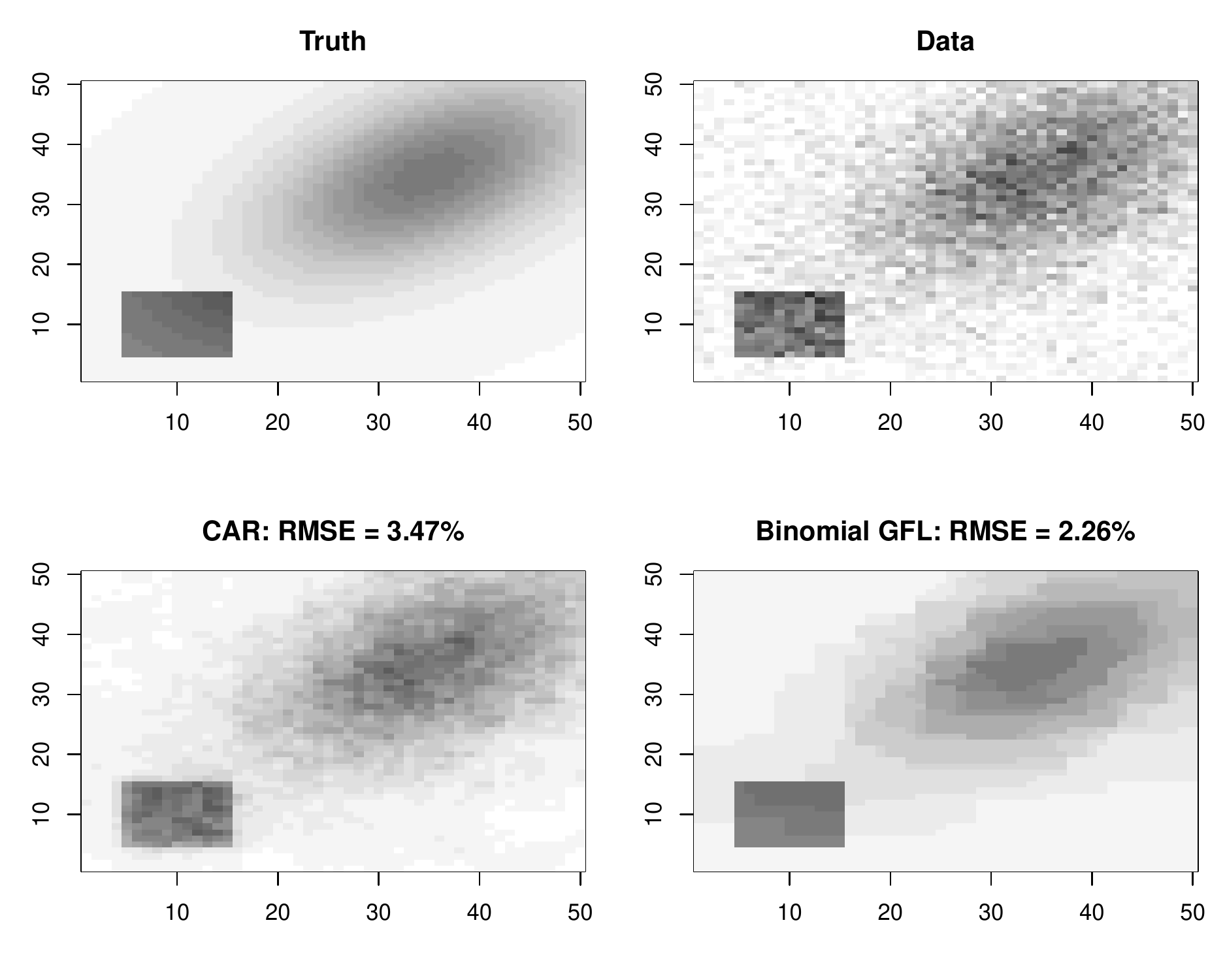}
\end{subfigure}
\caption{\label{fig:gfl_versus_car} The binomial graph-fused lasso (GFL) versus the posterior mean under a Gaussian intrinsic auto-regressive (CAR) model on two examples: smooth (top four panels) and discontinuous (bottom four panels).  The panels labeled ``Truth'' show $w^{(s)}$, while the panels labeled ``Data" show the raw frequencies $y^{(s)}/m^{(s)}$.  (Black is 1, white is 0.)  The GFL solution has slightly worse performance (as measured by root mean-squared error) on the smooth problem, but much better performance on the problem with a discontinuity.}
\end{figure}

Figure \ref{fig:gfl_versus_car} provides some intuition on a toy problem.  The top four panels show the binomial graph-fused lasso versus the posterior mean from a Gaussian intrinsic autoregressive model on a problem where the true signal is smooth.  The bottom four panels show the same comparison for a problem where the true signal has a discontinuity.  For the smooth problem, the $\ell_1$ MAP estimate has only slightly worse root mean-squared error than the CAR model, while for the problem with the discontinuity, it has a much better RMSE.  Our experiments later in the paper elaborate upon this point in more detail: the loss of using an $\ell^1$ penalty on a smooth signal are small, compared to the gains of using it when the signal has regions of varying smoothness (including discontinuities).

We recognize that convex penalties come with their own limitations, in that they have a tendency to overshrink large signals (in this case, large jumps across edges in the graph) in the presence of sparsity.  See \citet{fan:li:2001} and \citet{Carvalho:Polson:Scott:2008a} for frequentist and Bayesian perspectives on this issue, respectively, as well as the earlier paper by \citet{geman:reynolds:1992} for a discussion of the same issue in image processing.  Incorporating nonconvex penalties into spatial smoothing is an active area of research, and we describe one such approach in Section \ref{sec:bayesian_smoothing}.  But this requires MCMC, which unfortunately does not scale as well as our method.


\subsection{Algorithmic details}
\label{sec:algorithm_details}

In order to solve \eqref{eqn:logit_objective0} at scale, we use a variant of a recent method for solving the graph-fused lasso problem based on trail decompositions \citep{tansey:scott:2015}. This algorithm leverages a fundamental result in graph theory which states that all connected graphs with $2k$ odd-degree vertices can be decomposed into a set of $k$ non-overlapping trails. This enables us to rewrite the penalty term in \eqref{eqn:logit_objective0} as a summation over trails $\mathcal{T}$:
\begin{equation}
\label{eqn:logit_objective1}
\begin{aligned}
& \underset{\beta \in \R^p}{\text{minimize}}
& & 
\sum_{s \in \mathcal{V}} \left[ m^{(s)} \log \left\{ 1+ \exp (\beta^{(s)}) \right\} - y^{(s)} \beta^{(s)} \right ] + \lambda \sum_{\tau \in \mathcal{T}} \sum_{(r,s) \in \tau} |\beta^{(r)} - \beta^{(s)}| \, .
\end{aligned}
\end{equation}
 Denoting the negative log likelihood as $\ell(\mathbf{y}, \boldsymbol\beta)$, slack variables $\mathbf{z}$ are then introduced for each $\beta^{(i)}$ in the penalty, which results in the following equivalent problem:
\begin{equation}
\label{eqn:logit_objective2}
\begin{aligned}
& \underset{\boldsymbol\beta \in \R^n}{\text{minimize}}
& & 
\ell(\mathbf{y}, \boldsymbol\beta) + \lambda \sum_{\tau \in \mathcal{T}} \sum_{(r,s) \in \tau} |z^{(r)} - z^{(s)}|\, . \\
& \text{subject to}
& & \beta^{(r)} = z^{(r)} \\
& & & \beta^{(s)} = z^{(s)} \, ,
\end{aligned}
\end{equation}
where the constraints hold for all pairs $(r,s) \in \tau$, for all $\tau \in \mathcal{T}$.  We then solve \eqref{eqn:logit_objective2} via the Alternating Direction Method of Multipliers, or ADMM \citep{boyd:etal:2011}.

Because $\ell(\mathbf{y}, \boldsymbol\beta)$ is not squared-error loss, a small modification of the algorithm in \citet{tansey:scott:2015} is required.  As in Newton's method, at every step of the algorithm we iteratively form a second-order Taylor expansion to $\ell(\mathbf{y}, \boldsymbol\beta)$ at the current value of $\boldsymbol\beta$, and use this as a surrogate loss function $\tilde\ell(\tilde{y}, \boldsymbol\beta)$.  This yields the following update steps:
\begin{align}
\label{eqn:admm_updates_p}
p^{(s)} & \coloneqq \frac{1}{1 + e^{-\beta^{(s)}}} \\
\label{eqn:admm_updates_w}
\omega^{(s)} & \coloneqq m^{(s)} p^{(s)} (1 - p^{(s)}) \\
\label{eqn:admm_updates_y}
\tilde{y}^{(s)} & \coloneqq \beta^{(s)} - \frac{m^{(s)} p^{(s)} - y^{(s)}}{\omega^{(s)}} \\
\label{eqn:admm_updates_beta}
\beta^{(s)} & \coloneqq \frac{2\tilde{y}^{(s)}\omega^{(s)} + \alpha \sum_{j \in \mathcal{J}^{(s)}} (z^{(j)} - u^{(j)})}{2\omega^{(s)} + \alpha |\mathcal{J}^{(s)}|} \\
\label{eqn:admm_updates_z}
\mathbf{z}^{(\tau)} & \coloneqq \underset{\mathbf{z}}{\text{argmin}} \left( \frac{\alpha}{2} \sum_{s \in \tau} (\beta^{(s)} + u^{(s)} - z^{(s)})^2 + \lambda \sum_{(r,s) \in \tau} |z^{(r)} - z^{(s)}|  \right) \; , \quad \forall \; \tau \in \mathcal{T} \\
\label{eqn:admm_updates_u}
\mathbf{u} & \coloneqq \mathbf{u} + A\boldsymbol\beta - \mathbf{z} \, ,
\end{align}
where $\mathbf{z}^{(\tau)}$ is the set of slack variables in the $\tau^{\text{th}}$ trail, $u$ is the scaled ADMM dual variable, $\alpha$ is the scalar ADMM step-size parameter, $\mathcal{J}^{(s)}$ is the set of slack-variable indices that map to $\beta^{(s)}$, and $A$ is a sparse binary matrix used to encode the appropriate $\beta^{(s)}$ for each $z^{(j)}$.  We iterate the updates in \eqref{eqn:admm_updates_p}-\eqref{eqn:admm_updates_u} in order until the dual and primal residuals have sufficiently small norms.  Note that the update in \eqref{eqn:admm_updates_z} requires solving a weighted 1-dimensional fused lasso problem.  This can be done in linear time via an efficient dynamic-programming routine \citep{johnson:2013,glmgen:2014}. 

We select the $\lambda$ hyperparameter separately for each nonterminal node of the dyadic density decomposition, as the level of smoothing required at various scales may differ. A common approach to tuning hyperparameters is to use cross-validation. However, it is not clear in our case how to split our data into training and testing splits such that cross-validation would make sense. Furthermore, since we are using a convex optimization routine, we can use warm starts to quickly explore the solution path over a grid of $\lambda$ values.  We then compare the fits for different $\lambda$'s using an information criterion, leveraging the results on the degrees of freedom for $\ell_1$-based graph smoothers in \citet{wang2014trend}.  We use BIC for our selection criterion, as it is more conservative in finding changepoints than AIC or AICc, although results are similar across all criteria.

\subsection{A Bayesian variation}
\label{sec:bayesian_smoothing}

In this section, we describe a Bayesian variation of our estimator.  This approach can be used on problems of small to moderate size, and offers three advantages over the classical formulation: 1) It allows for uncertainty quantification; 2) It offers a more principled approach to choosing the hyperparameter $\lambda$; and 3) it offers the chance to move beyond the $\ell^1$ penalty, via the choice of heavy tailed priors that preserve large jumps in the signal across edges in the graph.  The tradeoff is that it requires MCMC, leading to much longer computation times.

We first observe that the graph-fused lasso estimator presented above is a special case of a maximum a posteriori (MAP) estimate of $\beta$ under the unnormalized posterior
\begin{equation}
\label{eqn:gtf_laplace_gamma_posterior}
\begin{aligned}
P(\boldsymbol\beta \mid \mathbf{y}) \propto  \exp\left\{-\ell(\mathbf{y}, \boldsymbol\beta) \right\} \cdot \exp \left\{ - \lambda \vnorm{\Delta^{(K+1)}\boldsymbol\beta}_1\right\} \, ,
\end{aligned}
\end{equation}
where $\Delta^{(K+1)}$ is the order-$K$ graph trend filtering penalty matrix \citep{wang2014trend}.   Concretely, when $K=0$, the matrix $\Delta^{(1)}$ is the oriented edge matrix for the graph $\mathcal{G}$, which has one row for every edge $e$, and for which
$$
\Delta^{(1)}_{e, r} = 1 \, , \quad \Delta^{(1)}_{e, s} = -1 \, , \quad \mbox{and }\Delta^{(1)}_{e, t} = 0 \mbox{ for } t \notin \{s,r\} \, .
$$
This matrix extracts the first-order differences in $\bbeta$ across edges, recovering the penalty in (\ref{eqn:logit_objective0}).  Higher-order penalty matrices ($K>1$) are defined recursively, as in \citet{wang2014trend}.  Throughout, we use $d$ to denote the number of rows in $\Delta^{(K+1)}$.

In (\ref{eqn:gtf_laplace_gamma_posterior}), the prior over $\bbeta$ is a product of independent Laplace priors over the the $(K+1)$-order differences in $\bbeta$ across edges.  This prior can be represented as a scale mixture of normals.  Letting $\Delta \equiv \Delta^{(K+1)}$, we have
\begin{align}
p(\beta \mid \lambda^2, \Omega) &\propto \exp \left\{ - \frac{\lambda^2}{2} \bbeta^T (\Delta^T \Omega^{-1} \Delta) \bbeta \right\}  \label{eqn:conditional_gaussian} \\
\Omega^{-1} &= \text{diag}(\omega_1^{-1}, \ldots, \omega_d^{-1}) \, , \quad \, \omega_j \sim p(\omega) \, .
\end{align}
If $\omega^{-1}$ has an exponential distribution, the Laplace prior in (\ref{eqn:gtf_laplace_gamma_posterior}) is recovered marginally.  But if $p(\omega)$ is some other mixing distribution, we get some other global/local shrinkage prior on each entry in $\Delta^{(K+1)}\boldsymbol\beta$.  \citet{roualdes:2015} and \citet{faulkner:minin:2015} explore this framework for the one-dimensional trend-filtering case, i.e.~where $\mathcal{G}$ is a chain graph, assuming Gaussian error.  We explore this same class of priors for binomial models on general graphs, in the context of the density smoothing problem.

For the case $K=0$, which penalizes first differences, there is a close connection with Markov random fields, like the kind used in \citet{zhou:hanson:2011}.  Because each diagonal entry in $\Omega$ corresponds to an edge $(r,s)$ of the graph,  we can rewrite (\ref{eqn:conditional_gaussian}) as
$$
p(\beta \mid \lambda^2, \Omega) \propto \exp \left\{ - \frac{\lambda^2}{2} \sum_{(r,s) \in \mathcal{E}} \omega_{rs}^{-1} (\beta^{(r)} - \beta^{(s)})^2 \right\}   \, ,
$$
which, conditionally on $\lambda$ and $\Omega$, is a Gaussian Markov random field.  This prior is improper on $\mathcal{R}^p$ because $\Delta^T \Omega \Delta$ is singular, but is proper on a subspace defined by an appropriate constraint.  Specifically, restricting inference to the subspace implied by the centering constraint $\{ \bbeta \in \mathcal{R}^p : \mathbf{1}^T \bbeta = 0 \}$, where $\mathbf{1}$ is the vector of all 1's, ensures propriety \citep[c.f.][Ch.~3.3]{banerjee:etal:2004}.   This follows from the fact that $\Delta^T \Omega \Delta$ is a weighted graph Laplacian matrix, whose nullspace is spanned by $\mathbf{1}$ if all diagonal entries in $\Omega$ are positive.  Analagous constraints ensure propriety for higher-order penalty matrices, when $K > 1$.

\paragraph{Details.}  For the mixing distribution $p(\omega)$, we use the choice leading to the generalized double Pareto prior in \citet{armagan:etal:2013}, also called the gamma-lasso prior described in \citet{taddy:2010}.  This choice leads to heavy tails, thereby avoiding some of the overshrinkage problems of the $\ell^1$/Laplace prior.  Many such heavy-tailed shrikage priors have been proposed in the literature; see \cite{Polson:Scott:2010a} for a review.  We use the gamma-lasso because it also has a simple interpretation as a mixture of Laplace priors, making the connection with the graph fused lasso explicit, and because our experiments with other choices led to essentially equivalent performance.  We represent the binomial likelihood using the P\'olya-Gamma data augmentation scheme described in \citet{polson:scott:windle:2012a}.  This yields conditionally conjugate MCMC updates for all model parameters.

To enforce the propriety constraint, we add in a global intercept term to the binomial sampling model, i.e.~
$$
(y_{\gamma0^{(s)}} \mid y_{\gamma}^{(s)}) \sim \mbox{Binomial}(y_{\gamma}, w_{\gamma}) \, , \quad
w^{(s)} = \frac{ \exp\{ \alpha + \beta^{(s)} \} } {1 + \exp\{ \alpha + \beta^{(s)} \} } \, .
$$
The intercept $\alpha$ is assigned a vague prior, and we center $\beta^{(s)}$'s at every MCMC iteration.

The natural full Bayes approach to fitting this model is to include the global shrinkage parameter $\lambda^2$ in the sampler.  In doing so, we encountered severe problems, both with poor mixing (a widely known issue with global/local shrinkage priors) and with numerical stability of matrix operations.  We were unable to get either of the MCMC schemes described for the 1D trend-filtering model by \citet{roualdes:2015} and \citet{faulkner:minin:2015}---Gibbs sampling and Hamiltonian Monte Carlo, respectively---to yield stable performance for general graphs.  These issues were addressed by fitting separate models across a discrete grid of fixed values for $\lambda^2$, and choosing $\lambda^2$ using the deviance information criterion \citep{spiegelhalter:etal:2002}.  Thus our approach is empirical Bayes, rather than fully Bayes.  Full details can be found in the appendix.

\section{Simulations}
\label{sec:simulations}

\subsection{Spatially varying Gaussians}

\begin{figure}
\centering
\begin{subfigure}[b]{\textwidth}
\begin{tabular}{ccc}
\hspace{-0.3in}\raisebox{0.02\textheight}{\vspace{-0.1in}\includegraphics[height=0.175\textheight]{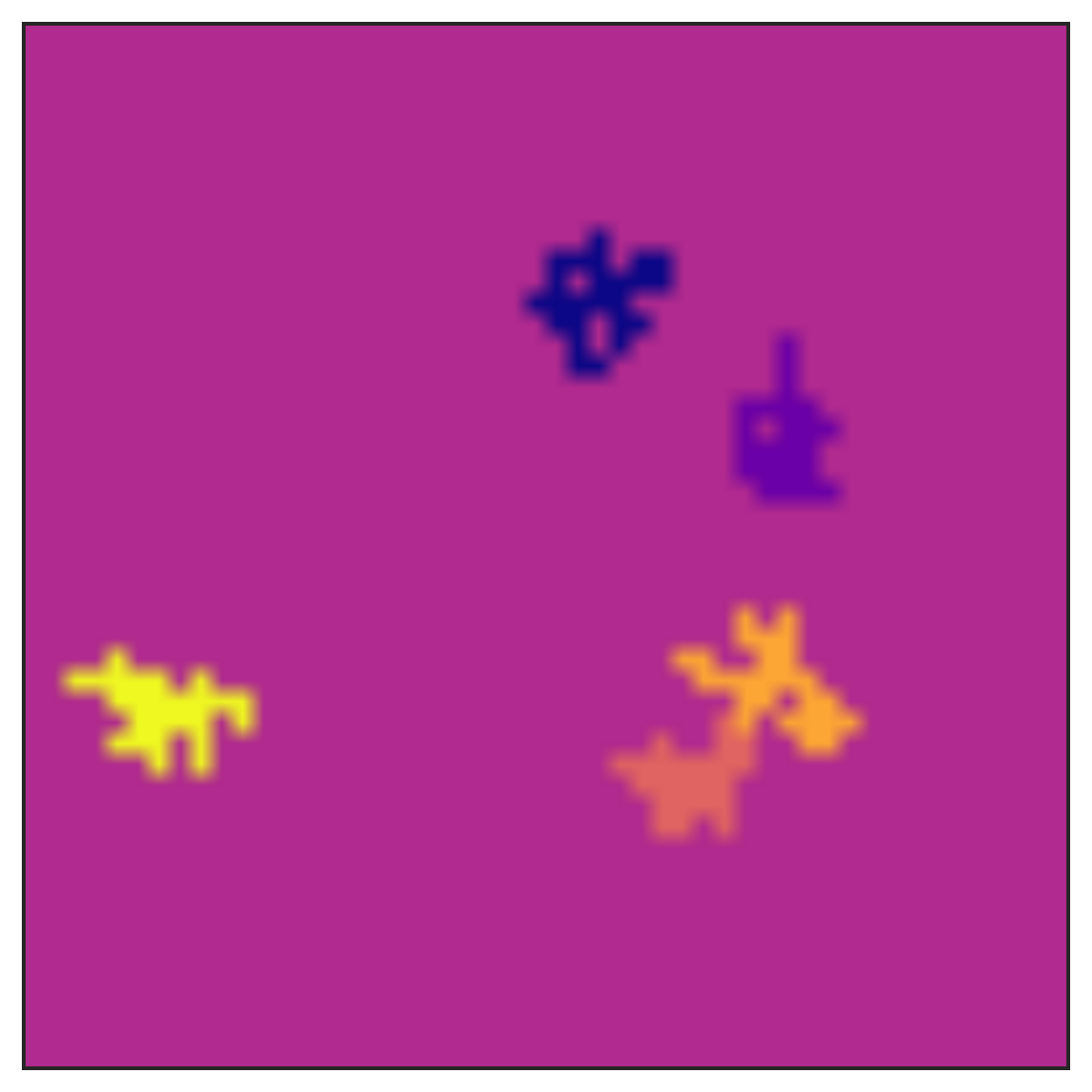}} &
\vspace{-0.1in}\includegraphics[width=0.27\textheight]{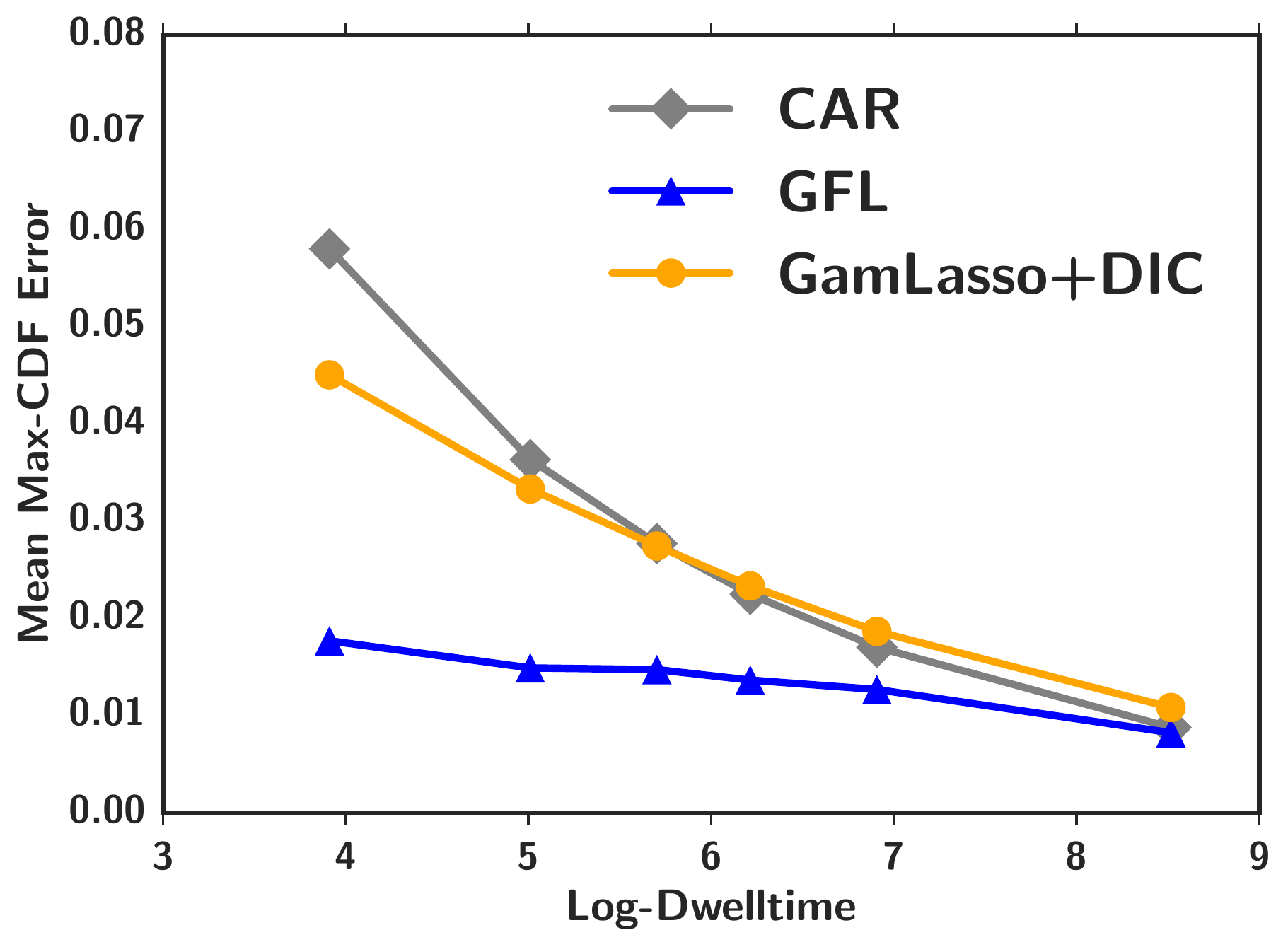} &
\raisebox{-0.01in}{\includegraphics[width=0.266\textheight]{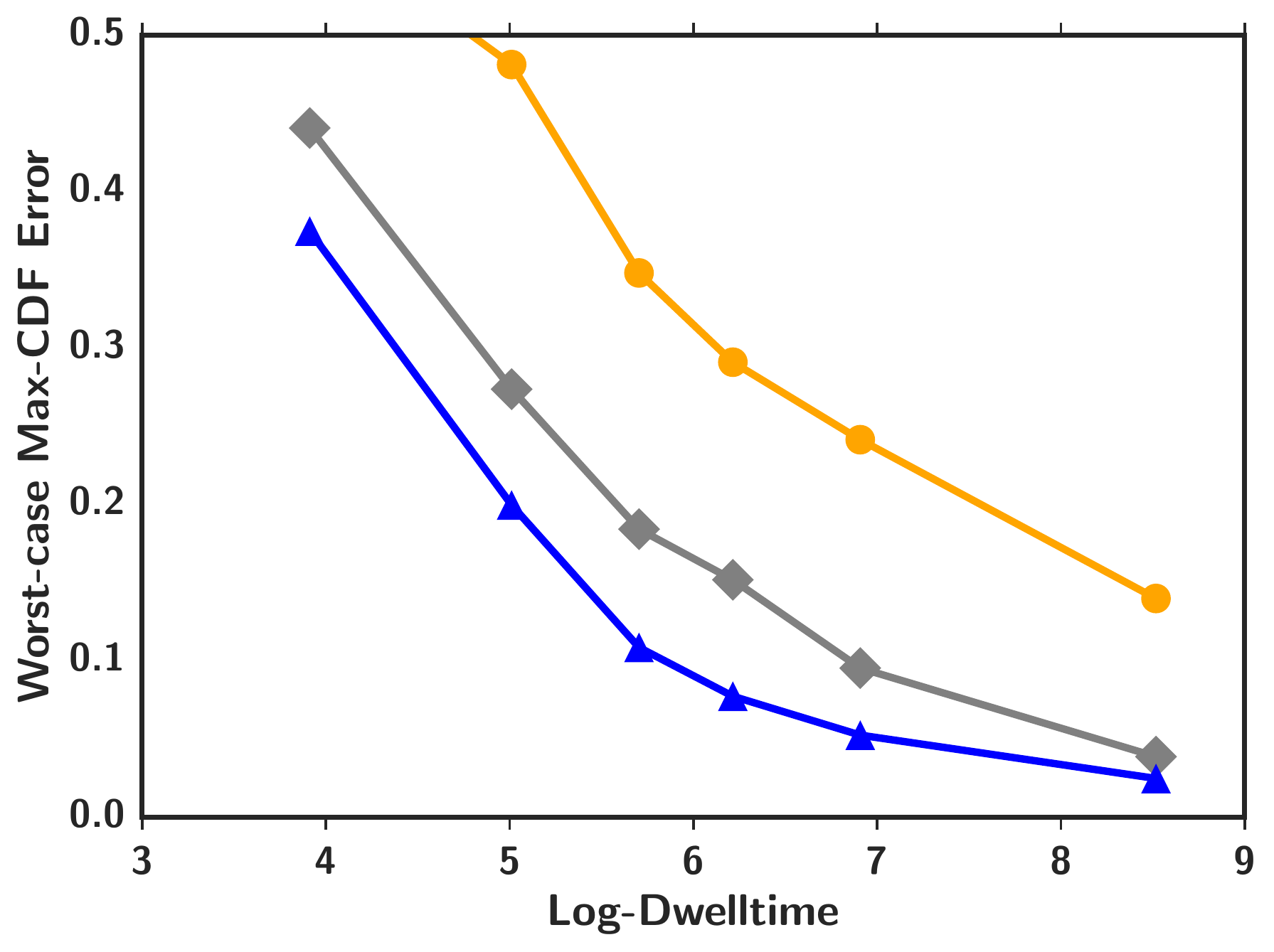}}
\end{tabular}
\label{fig:k0_means}
\caption{Piecewise constant}
\end{subfigure}

\begin{subfigure}[b]{\textwidth}
\begin{tabular}{ccc}
\hspace{-0.3in}\raisebox{0.02\textheight}{\vspace{-0.1in}\includegraphics[height=0.175\textheight]{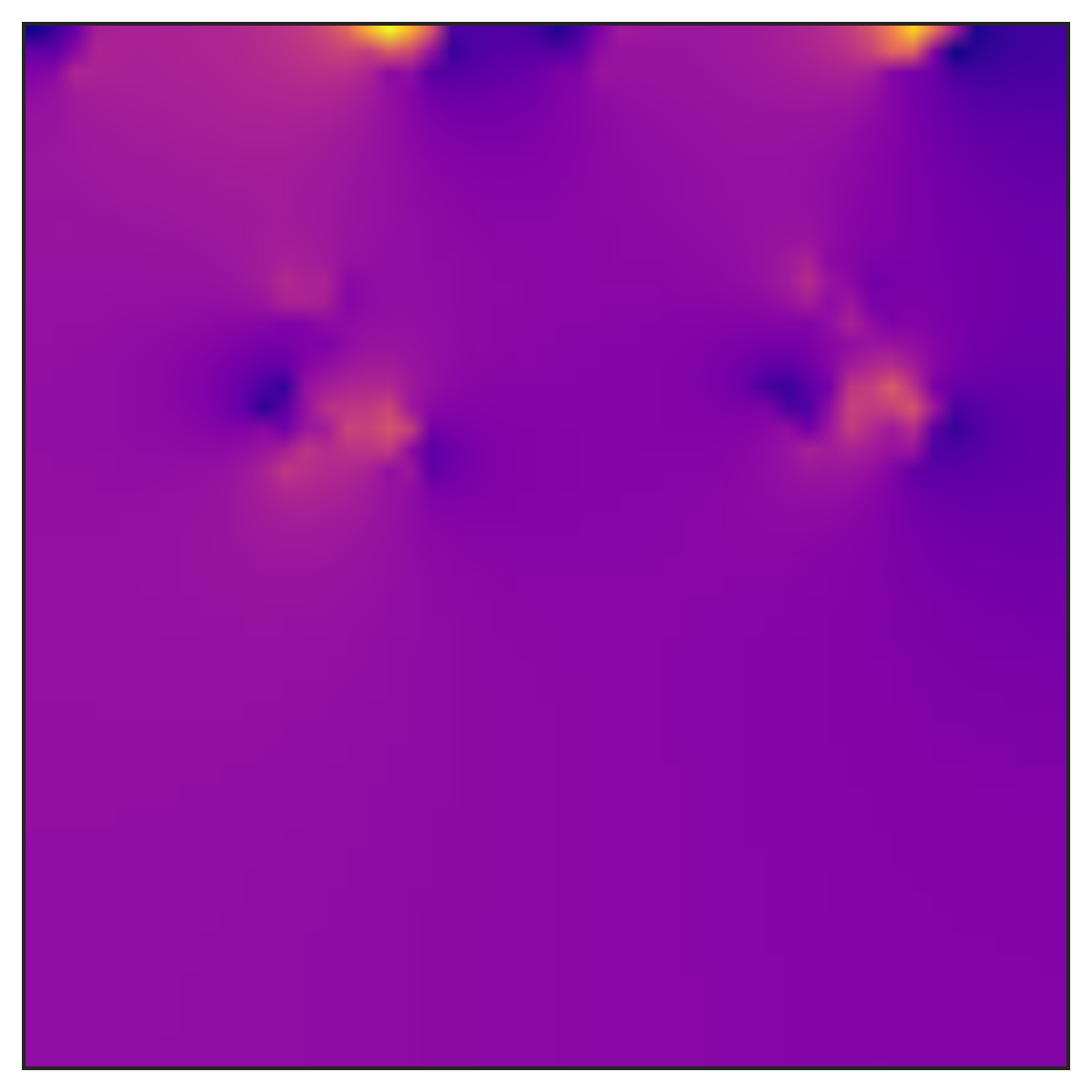}} &
\vspace{-0.1in}\includegraphics[width=0.27\textheight]{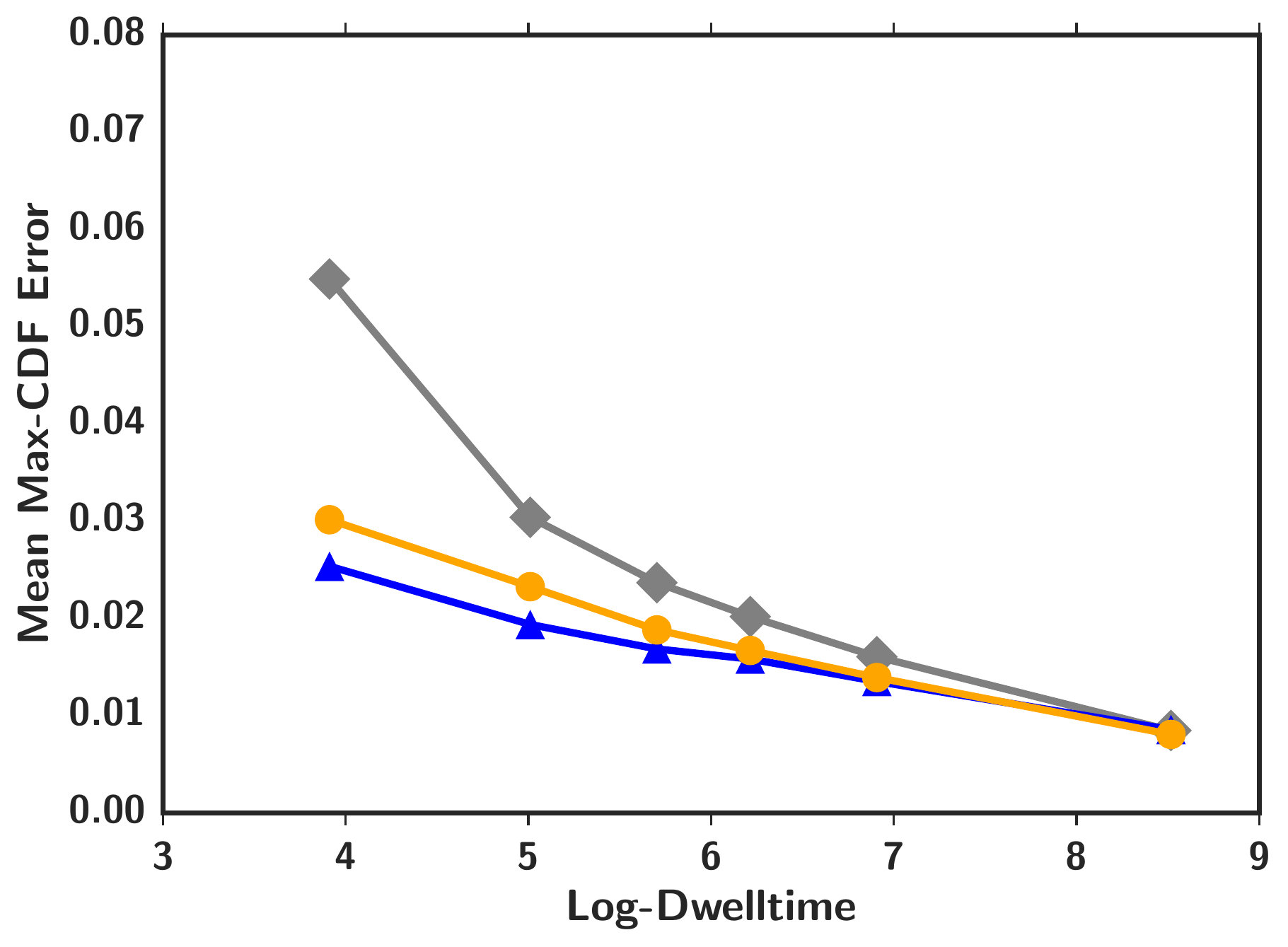} &
\raisebox{-0.01in}{\includegraphics[width=0.266\textheight]{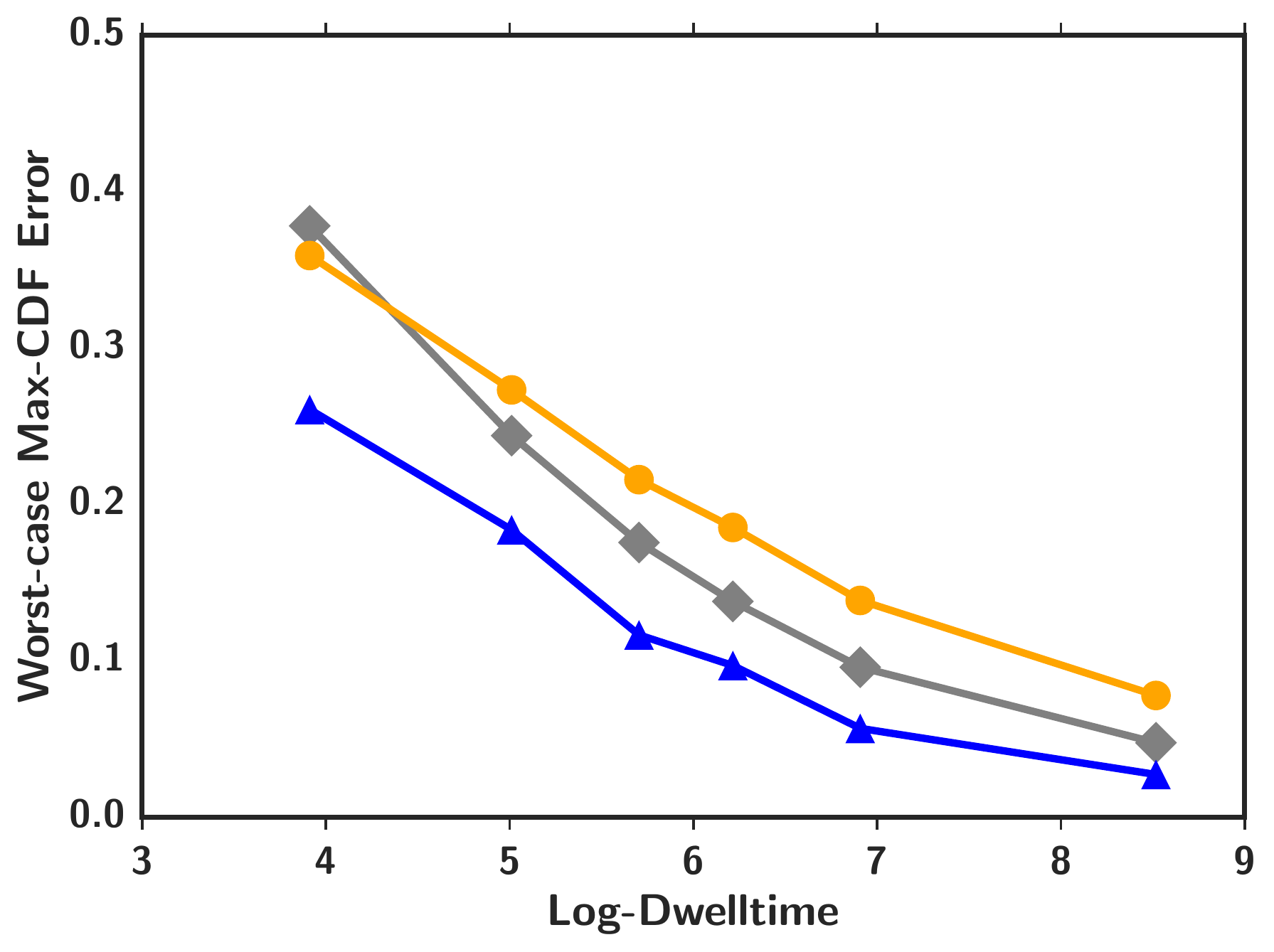}}
\end{tabular}
\label{fig:k1_means}
\caption{Piecewise linear}
\end{subfigure}

\begin{subfigure}[b]{\textwidth}
\begin{tabular}{ccc}
\hspace{-0.3in}\raisebox{0.02\textheight}{\vspace{-0.1in}\includegraphics[height=0.175\textheight]{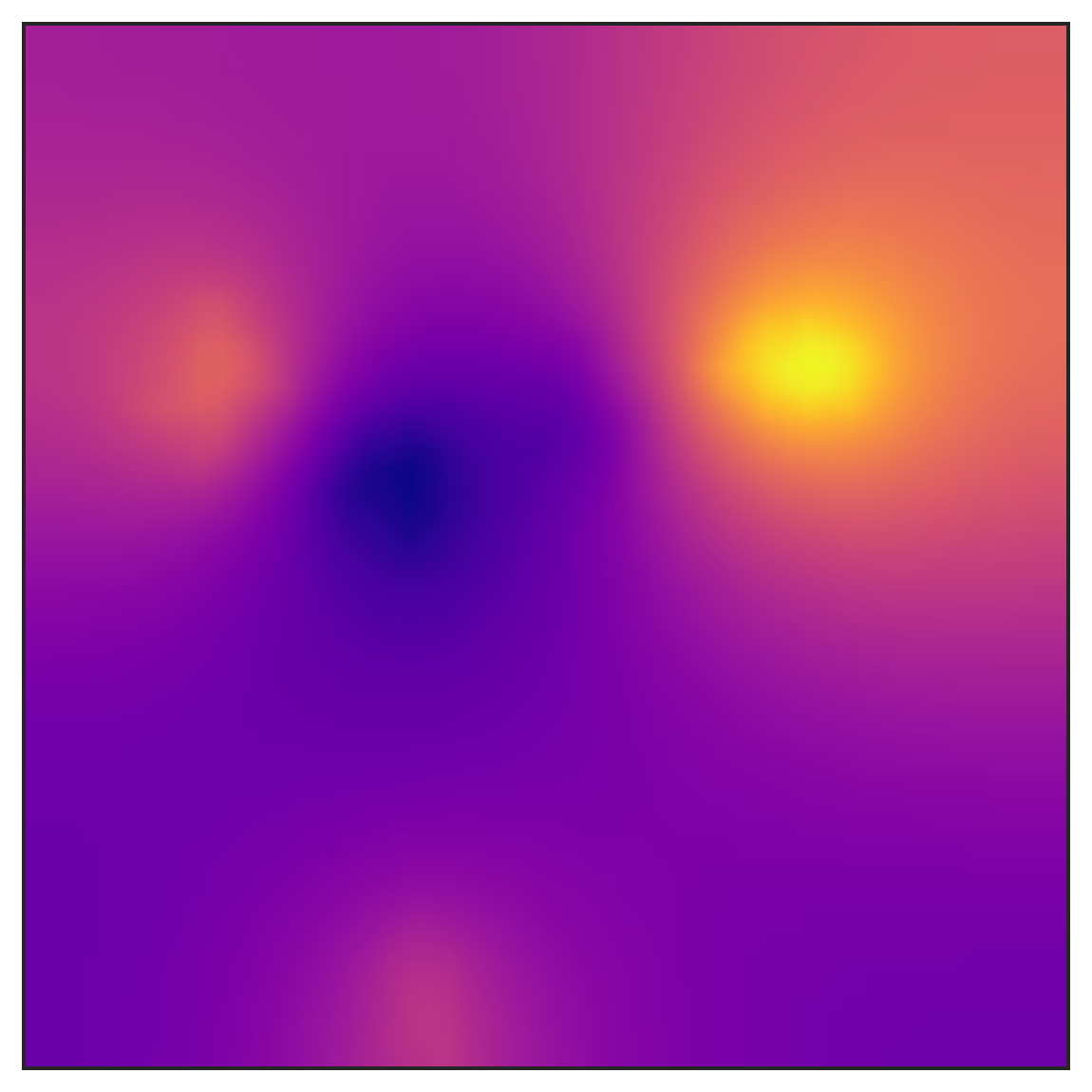}} &
\vspace{-0.1in}\includegraphics[width=0.27\textheight]{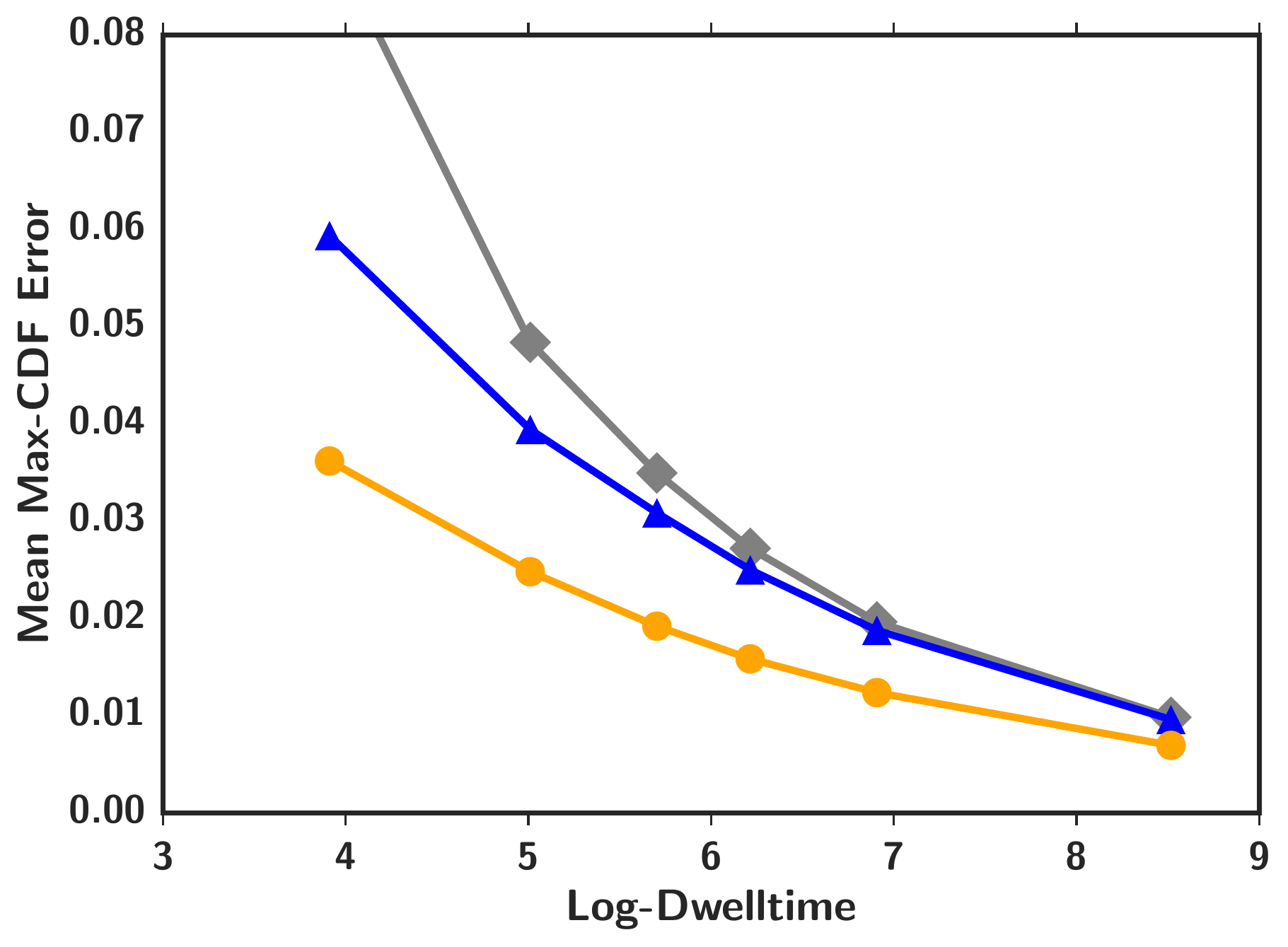} &
\raisebox{-0.01in}{\includegraphics[width=0.266\textheight]{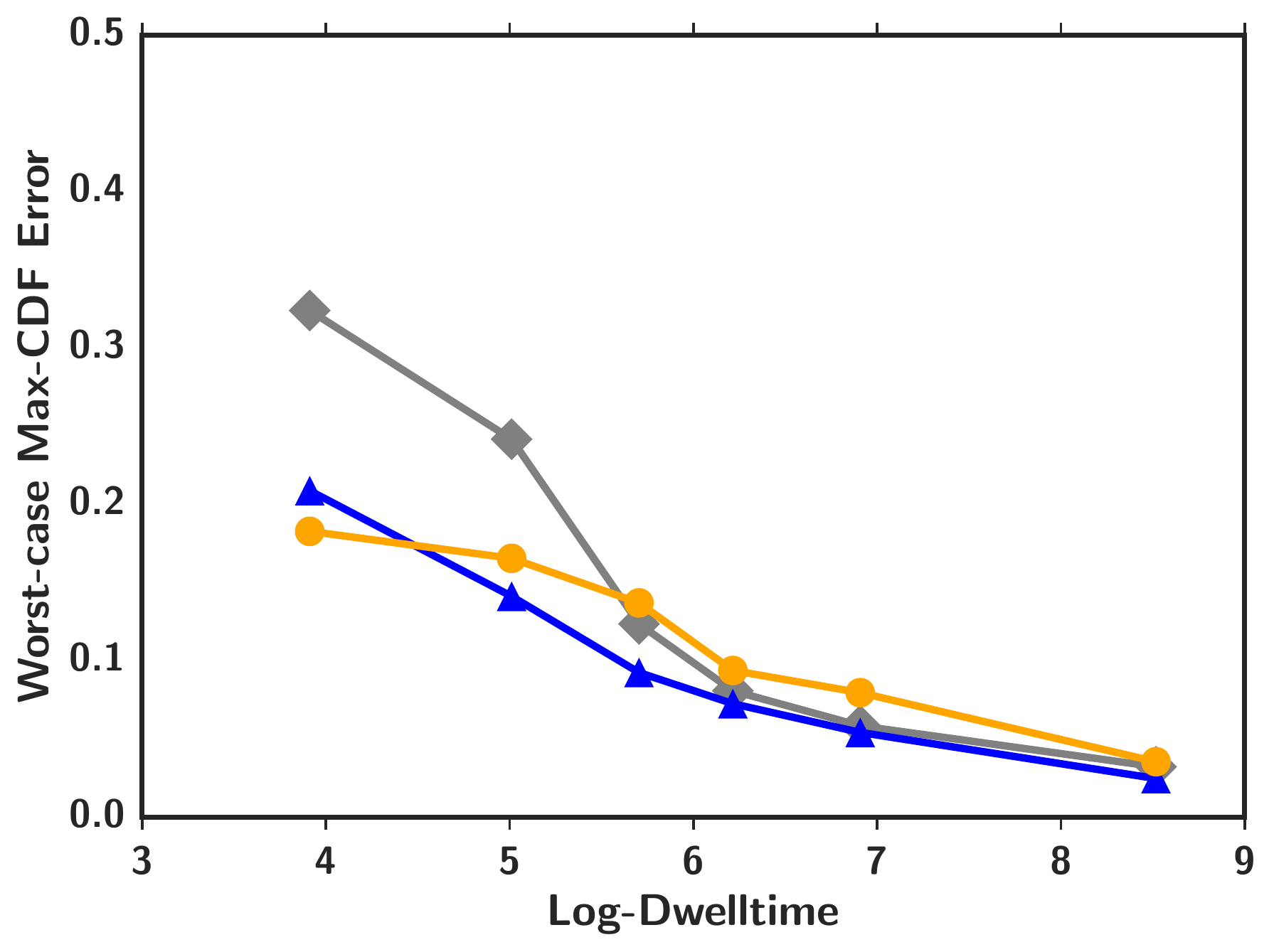}}
\end{tabular}
\label{fig:k2_means}
\caption{Piecewise quadratic}
\end{subfigure}

\begin{subfigure}[b]{\textwidth}
\begin{tabular}{ccc}
\hspace{-0.3in}\raisebox{0.02\textheight}{\vspace{-0.1in}\includegraphics[height=0.175\textheight]{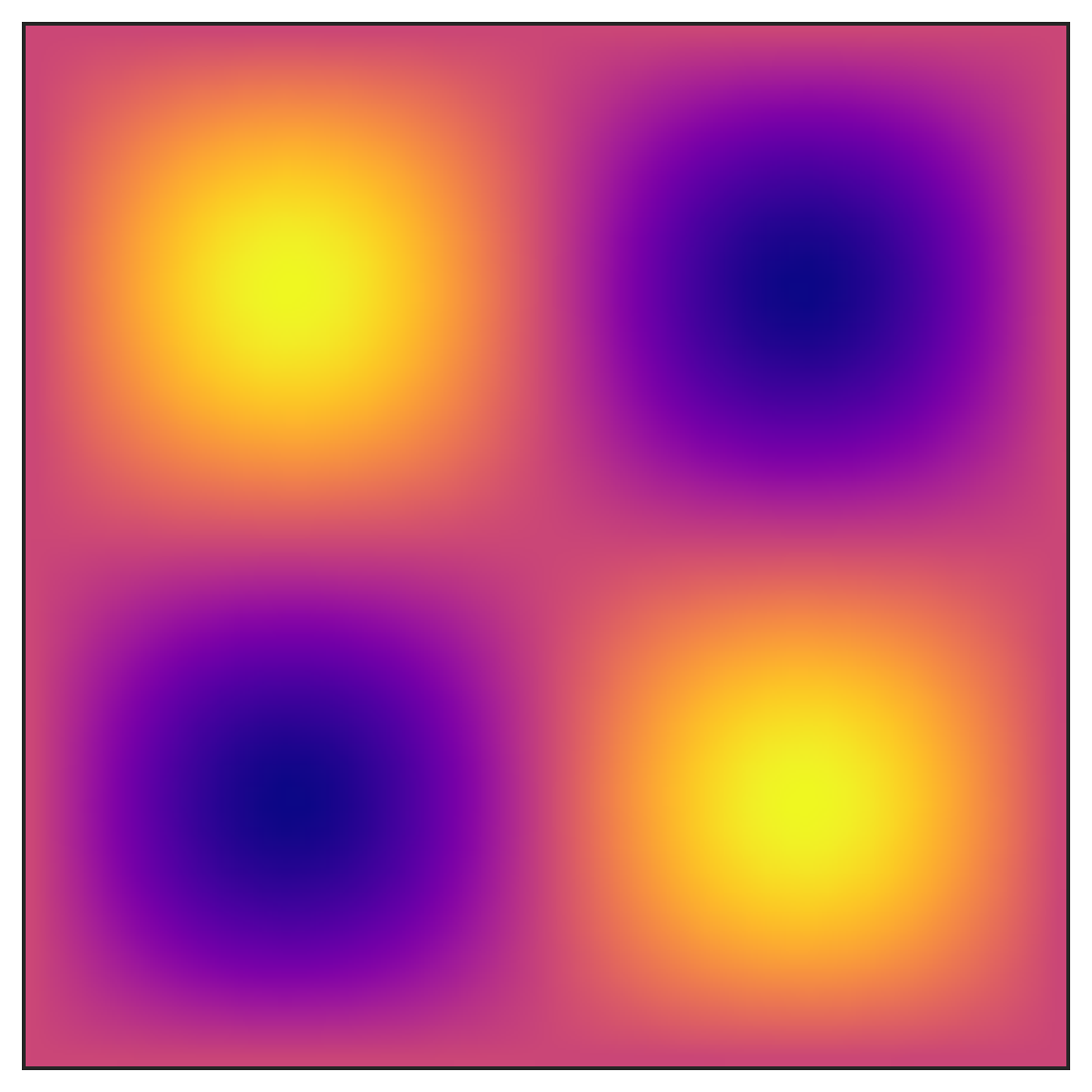}} &
\vspace{-0.1in}\includegraphics[width=0.27\textheight]{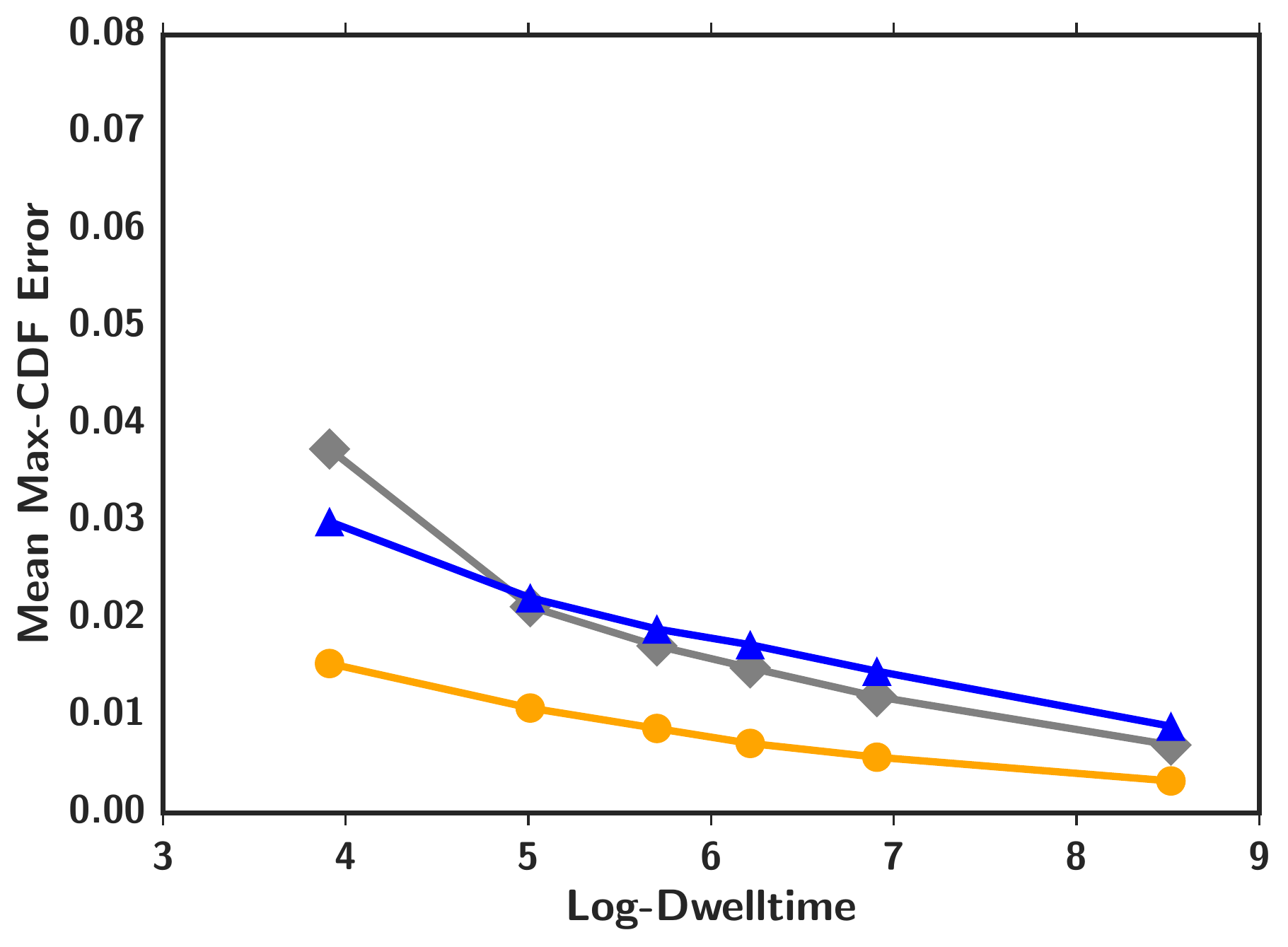} &
\raisebox{-0.01in}{\includegraphics[width=0.266\textheight]{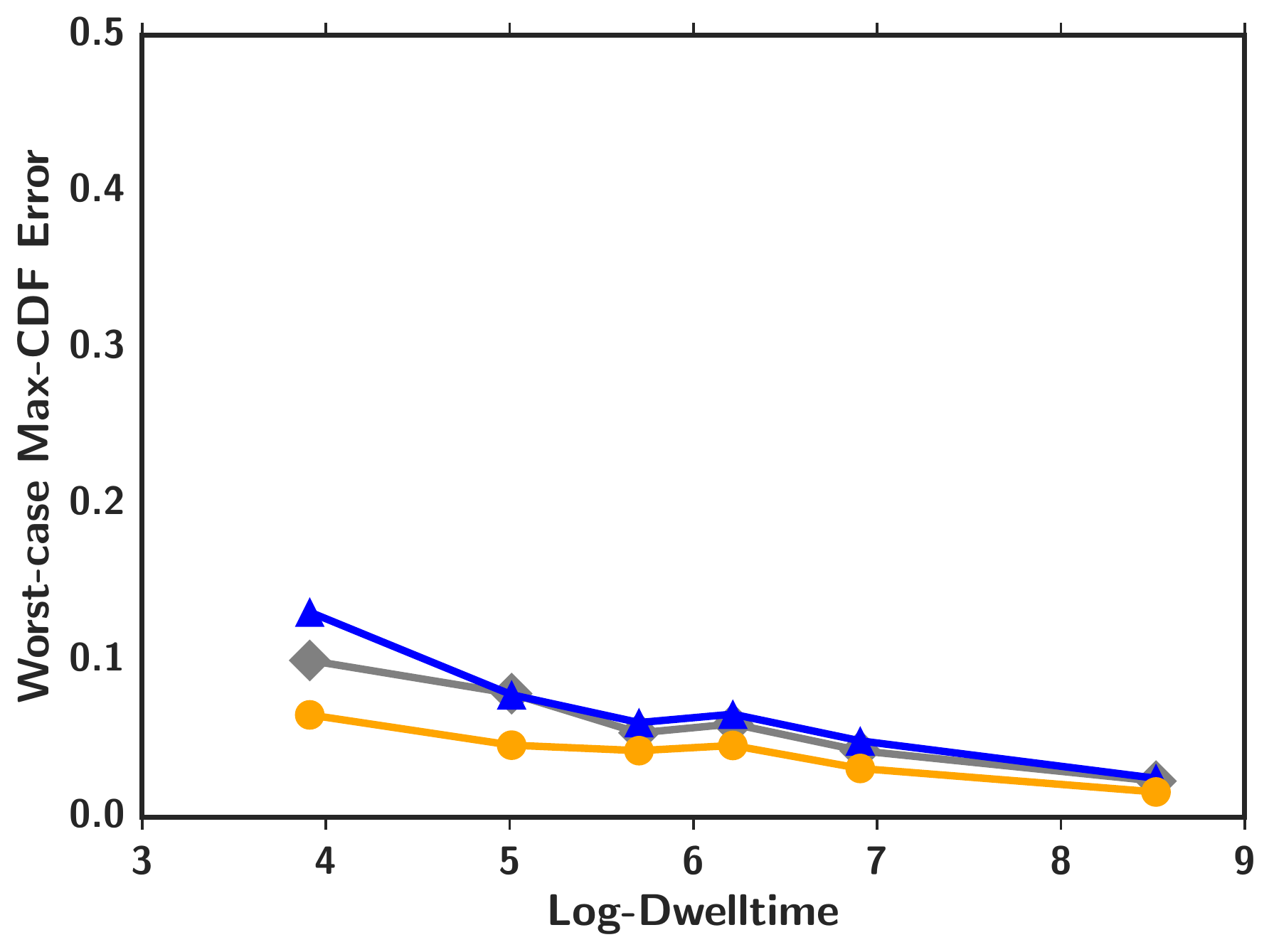}}
\end{tabular}
\label{fig:car_means}
\caption{Smooth}
\end{subfigure}

\caption{Results on Spatially-Varying Gaussian Benchmarks. The left column shows the spatially-varying means; the middle and right columns show the mean and worst-case performance of the three models across all spatial cells, respectively.}\label{fig:gaussian_means}
\end{figure}

We conducted a series of experiments to establish the performance of our proposed approach, including scenarios where the model is misspecified. In each experiment, we used a $50 \times 50$ grid, where the density $f^{(s)}$ in each grid cell was a Gaussian distribution with unit variance and a spatially-varying mean $\mu^{(s)}$ in the range $[-3, 3]$.  Such a problem is small enough for Bayesian approaches to be used as benchmarks.  We considered three spatial dependence scenarios for the mean based on piecewise polynomials, for which the vector $\Delta^{(K+1)} \mu$ had mostly zero entries: piecewise constant ($K=0$), linear $(K=1)$, and quadratic $(K=2)$.  These are very nearly splines of order 0 through 2, respectively; see \citet{wang2014trend}.   We also used a fourth, smooth signal having continuous derivatives of all orders, to which a Gaussian CAR model should be suited.  Figure \ref{fig:gaussian_means} (left column) shows the four different ground truths for the mean of each Gaussian in our experiments.  The observations in each grid cell were discretized into 2048 bins in the range $[-6,6]$.

We varied the number of observations per cell and measured the performance of each method, across all spatial cells. For the empirical Bayes method, we used a $K=2$ trend filtering model.  We also fit Bayesian CAR models by MCMC, following \citet{zhou:hanson:2011}. To measure the accuracy of the reconstruction, we computed the maximal CDF error between the true distribution at site $s$ and the estimated distribution:
$$
\mbox{distance}(F^{(s)}, \hat F^{(s)}) = \max_{x} \left| F^{(s)}(x)  - \hat F^{(s)}(x) \right| \, ,
$$
where $F$ and $\hat{F}$ are the true and estimated CDF, respectively.  The rationale for this error measure is that the ultimate use for the estimate is in a Kolmogorov--Smirnov test for anomalies, for which we need the CDF (see application in Section \ref{sec:main_application}).

Figure \ref{fig:gaussian_means} (middle and right columns) shows the results of these experiments.  For the piecewise constant and piecewise linear signals, the binomial graph-fused lasso (GFL) leads to the best average density reconstruction, beating both Bayesian models (second column in Figure).  For the piecewise quadratic and smooth signals, the binomial GFL and the Gaussian CAR model have comparable average performance, while empirical-Bayes trend filtering leads to the best estimate.

The third column of Figure \ref{fig:gaussian_means} also shows that, for all three piecewise polynomial signals, the GFL-based approach has the lowest worst-case error (where ``worst case'' means a method's highest CDF error across all spatial cells).  For the smooth signal, the GFL method has comparable worst-case error to the CAR model, and Bayesian graph trend-filtering using gamma-lasso priors has the best worst-case error.  This apparent near-minimaxity of the GFL method is important to the application in Section \ref{sec:main_application}, where a poor worst-case error can lead to inflated false-alarm rates for radiological anomaly detection.  We will also revisit this point in Section \ref{sec:large_sim}.

Overall, the empirical Bayes method seems strongest when the underlying variation is homogeneously smooth, while our proposed approach based on the graph-fused lasso yields performance that equals or betters a CAR model, but with far better scalability.  Since the empirical Bayes and CAR models require substantially more computation time, they are not likely to be practical for scaling to large datasets such as full-city radiological maps. For instance, in these synthetic experiments, estimating the spatial variation for a single split node in the binary tree takes 23.4 seconds for the CAR model on average (for 10,000 MCMC samples), versus 340 milliseconds for the GFL algorithm---nearly two orders of magnitude faster.  Moreover, each algorithm must be run hundreds of times, once for each splitting node in the tree.  Accordingly, the practical difference between the methods is one of getting a density estimate for all 2500 spatial cells in minutes, versus hours.  This disparity only widens as the size of the graph increases beyond the 50 $\times$ 50 case, and the task of verifying good MCMC mixing for the CAR and Bayesian GTF models becomes more difficult.

\subsection{Synthetic radiological data on a large grid}
\label{sec:large_sim}

We conducted a second simulation experiment to demonstrate the scalablity of our method, as well as to probe its advantages and disadvantages versus other scalable methods (in particular, Gaussian kernel smoothing).   As a ground-truth scenario, we created a $500 \times 500$ grid of density functions $f^{(s)}$, which is sufficiently large to capture fine-scale structure across the downtown area of a city.  We designed this scenario with two criteria in mind:
\begin{enumerate}
\item The densities $f^{(s)}$ should be nearly stationary (i.e.~all the same), so that kernel smoothing should lead to excellent performance for most spatial cells.  This presents a challenge to the density smoothing method we have proposed, which is not designed for stationary scenarios.
\item There should be a spatially isolated discontinuity, of the kind that might be caused by a line-of-sight occlusion or some other spatially localized radiological feature of the built environment.  This feature presents a localized challenge to Gaussian kernel smoothing, which assumes smooth variation.
\end{enumerate}

\begin{figure}[t]
\begin{center}
\includegraphics[width=6in]{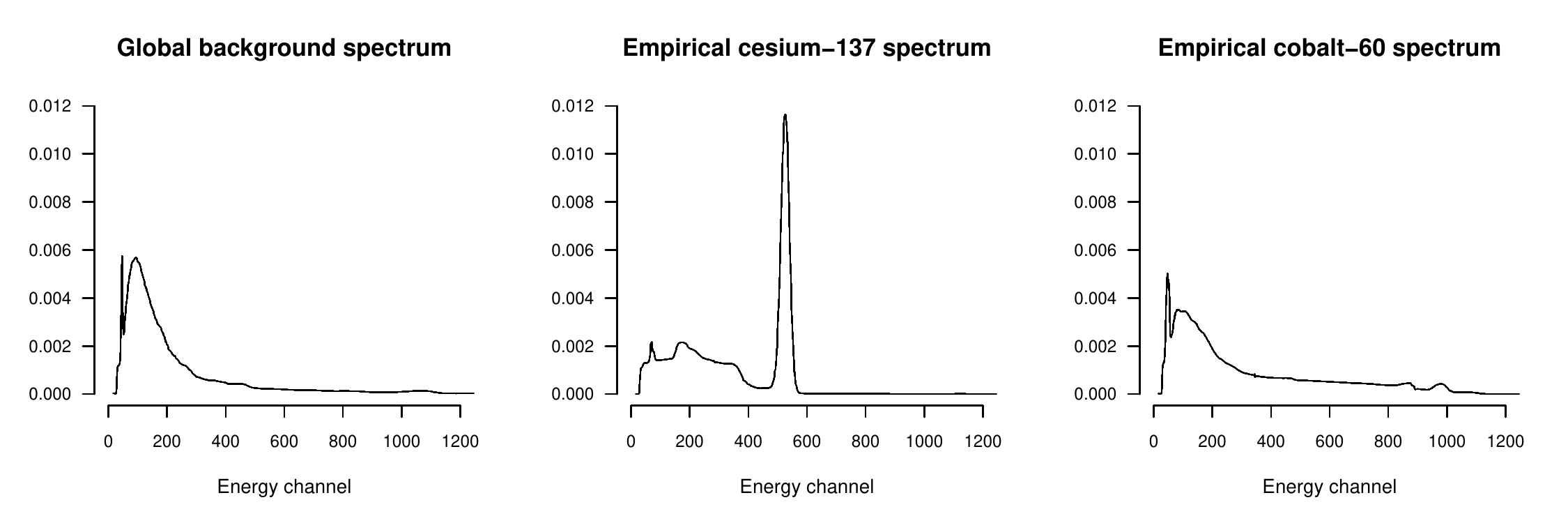}
\caption{\label{fig:bg_co_cs_densities} The gamma-ray spectral densities of the global background ($f_g$),  cesium-137 ($f_{\mathrm{Cs}}$), and cobalt-60 ($f_{\mathrm{Co}}$) used in our studies.}
\end{center}
\end{figure}

To meet these criteria, we constructed the density functions $f^{(s)}$ as a mixture of a global contribution $f_0$ plus site-specific deviations due to an isolated radiological source.  The global background density $f_0$ was chosen to be the empirically observed average background gamma-energy spectrum across all locations on the Pickle Research Campus, on the basis of 18 hours of data collection.  To create small deviations from stationarity, we then artificially injected two radioactive sources into the grid, at cell (101,101): (1) a 100-milliCurie source of cobalt-60, and (2) a 100-milliCurie source of cesium-137.

The spectral densities of the global background, cobalt-60, and cesium-137 are shown in Figure \ref{fig:bg_co_cs_densities}.  Note that these cesium and cobalt densities (denoted $f_{\mathrm{Co}}$ and $f_{\mathrm{Cs}}$, respectively) are not the known gamma-ray emission spectra of these two isotopes.  Rather, they are the empirical spectra estimated in a separate set of experiments involving the same detector used in our background-collection experiments.  The differences between the theoretical spectrum and the empirical spectrum of a radioactive isotope can be explained by many physical phenomena, such as Compton scattering.  Because our goal is to approximate what a real data-collection protocol would likely encounter, we use the empirical spectra rather than the theoretical spectra (see appendix).

\begin{figure}
\centering
\begin{subfigure}[b]{.7\textwidth}
\includegraphics[width=\textwidth]{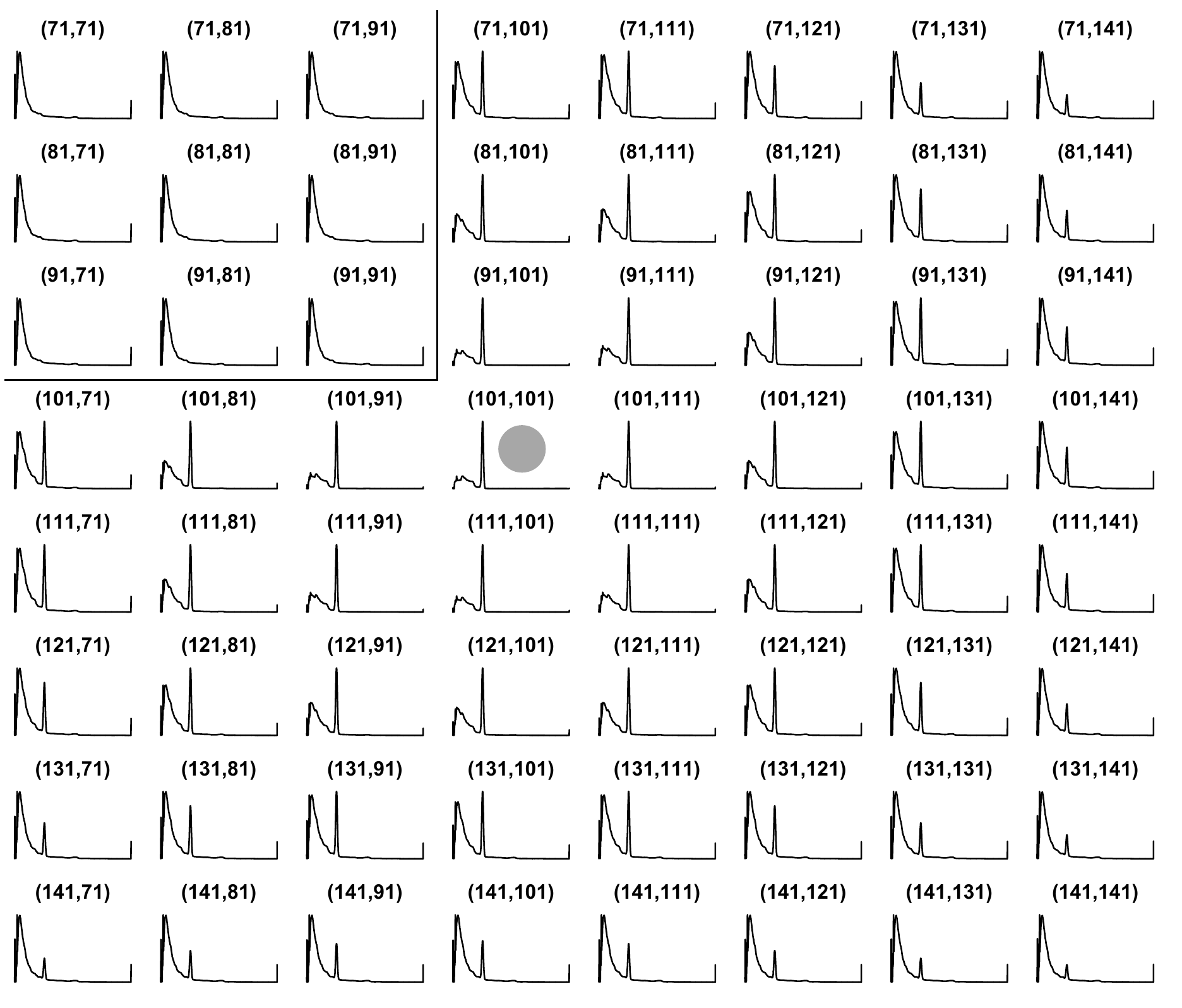}
\caption{Truth}
\end{subfigure}\qquad
\begin{subfigure}[b]{.2\textwidth}
\includegraphics[width=\textwidth,height=2.5cm]{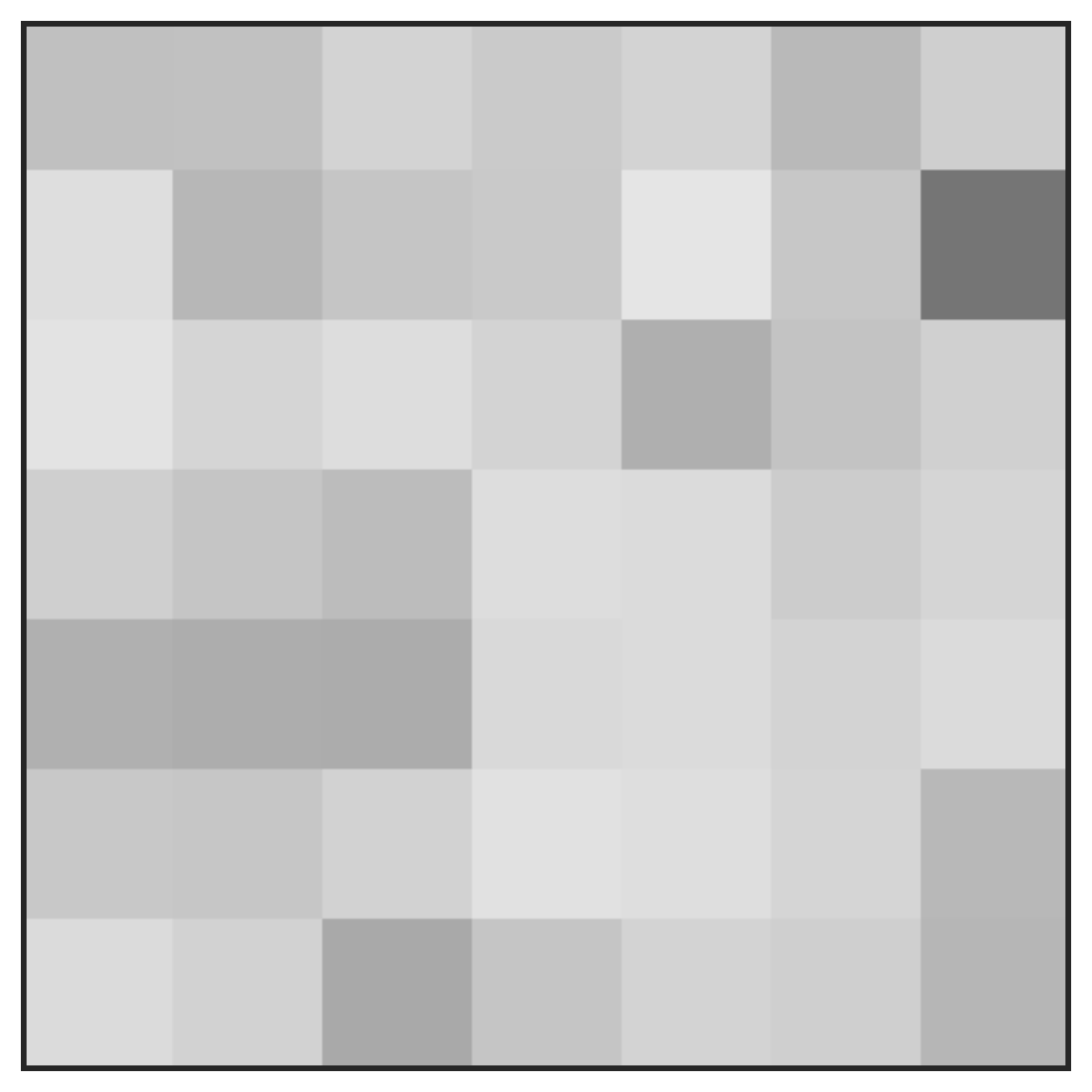}
\caption{Haar-Fisz error}

\vspace{2ex}

\includegraphics[width=\textwidth,height=2.5cm]{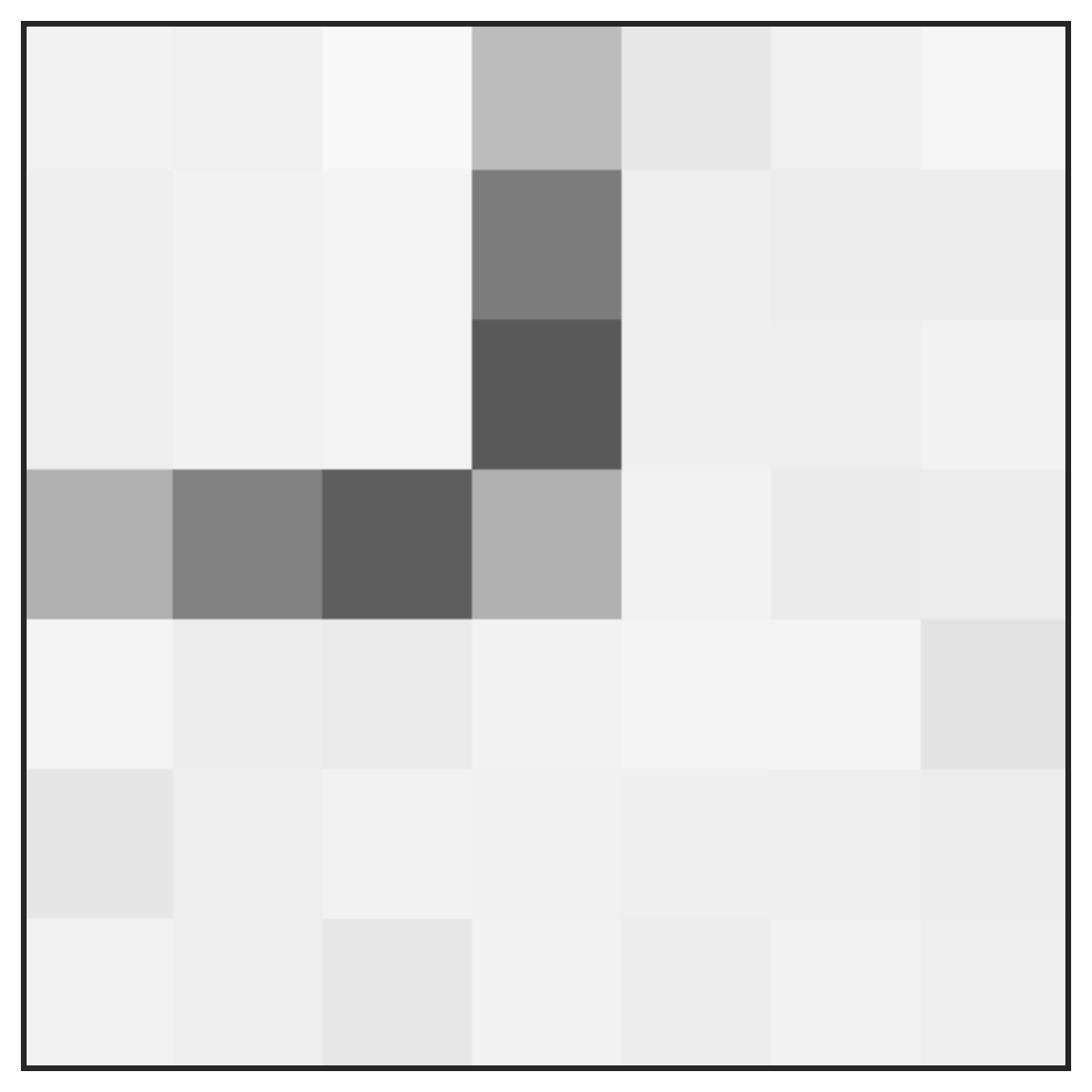}
\caption{Gaussian error}

\vspace{2ex}

\includegraphics[width=\textwidth,height=2.5cm]{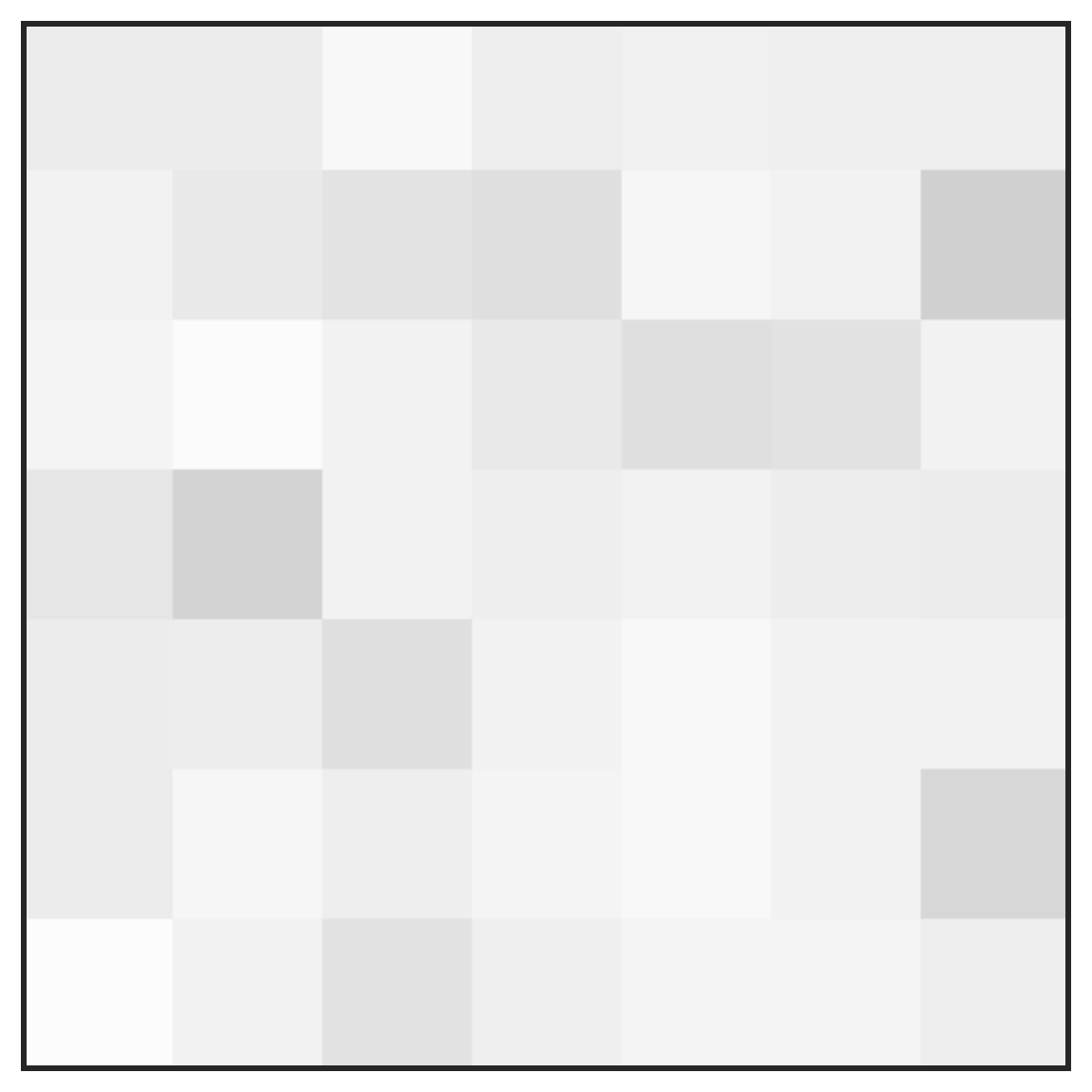}
\caption{GFL error}
\end{subfigure}
\caption{\label{fig:sim_study_corner} (Panel A) A subset of the true densities $f^{(s)}$ used in the large simulation study: every 10th row and column in the grid from cell (71,71) to cell (141, 141).  Cell (101,101) contains the isolated radiological source (grey dot).  To simulate a spatial discontinuity, there is a line-of-sight occlusion that prevents an observer at any site northwest of cell (101,101) from seeing emissions from this cesium source (black lines). (Panels B-D) Corresponding reconstruction errors for each method for the cells in Panel A. Both Haar-Fisz smoothing and the graph fused lasso method have errors that are nearly spatially invariant.  The Gaussian kernel smoother, despite its good average performance, has very poor performance where there is a discontinuity in the underlying spatial field. These spatially isolated errors will produce spurious anomalies when combined with the anomaly-detection technique described in Section 5.}
\end{figure}

Photons at each site were assumed to have energy drawn randomly from local density $f^{(s)}$, and to be detected at an average Poisson rate of $\lambda^{(s)} = \lambda_0 + \lambda^{(s)}_{\mathrm{Co}} +  \lambda^{(s)}_{\mathrm{Cs}}$.  Each term is defined as follows:
\begin{itemize}
\item $\lambda_0$ is the global background rate, assumed to be 40 photons per second in all cells, the average count rate observed during our field experiment.
\item $\lambda^{(s)}_{\mathrm{Co}}$ is the average rate of arrival of photons from the cobalt source at site $s$.
\item $\lambda^{(s)}_{\mathrm{Cs}}$ is the average rate of arrival of photons from the cesium source at site $s$.
\end{itemize}
To calculate $\lambda^{(s)}_{\mathrm{Co}}$ and $\lambda^{(s)}_{\mathrm{Co}}$ at each site, we used the following equation:
\begin{equation}
\label{eqn:sizetoCPS}
\lambda^{(s)}_{\mathrm{source}} = \frac{\mathrm{mCi}}{0.000844} \cdot 630 \cdot \left( \frac{0.05}{d_s} \right)^2 \cdot \exp \left\{  -0.0100029(d_s + 0.05) \right\} \, ,
\end{equation}
where $\mathrm{mCi}$ is the size of the source in milliCuries, and $d_s$ is the distance in meters from the center of grid cell $s$ to the source.  This equation is derived from well-understood physics, coupled with the field experiment described in the Appendix.

The ground-truth local densities were then specified via the appropriate site-specific convex combination of the background, cobalt, and cesium spectra:
$$
f^{(s)}(x) = \left( \frac{\lambda_0}{ K^{(s)} } \right) f_0(x)
+  \left( \frac{ \lambda^{(s)}_{\mathrm{Co}}  }{ K^{(s)} } \right) f^{(s)}_{\mathrm{Co}} (x) 
+  \left( \frac{ \lambda^{(s)}_{\mathrm{Cs}}  }{ K^{(s)} } \right) f^{(s)}_{\mathrm{Cs}} (x) \, ,
$$
where the common denominator $K^{(s)} = \lambda_0 +  \lambda^{(s)}_{\mathrm{Co}} + \lambda^{(s)}_{\mathrm{Cs}}$.  The exception is that, for all sites northwest of cell (101,101) in the grid, we simulated a line-of-sight occlusion, meaning that the cesium and cobalt photons do not reach these sites (i.e.~are scattered before they get there).  Thus $f^{(s)}(x) = f_0(x)$ for these cells.  This creates a discontinuity near cell (101,101).

It is difficult to visualize all 250,000 density functions at once, but Figure \ref{fig:sim_study_corner} shows them for a subset of 64 cells: every tenth row and column in the grid from cell (71,71) to cell (141,141).  Cell (101,101) contains the isolated radiological source.  It is easy to see the spatial decay in the contribution of this isolated source to the spectra of nearby cells, because cesium has a large peak in its spectrum around energy channel 550.  It is also easy to see the affect of the occlusion, in the 9 most northwesterly cells in the figure.

We emphasize that Figure \ref{fig:sim_study_corner} zooms in on the only region in the 500$\times$500 grid where there is any spatial variation.  The effect of the isolated source decays quite quickly as one moves away from this source; at cell (141,141), the cesium peak has nearly disappeared.  Everywhere else, the densities are nearly stationary.  This presents a very favorable scenario for Gaussian kernel smoothing (because linear shrinkage will be optimal across nearly the entire graph), and a challenge for our proposed method.

We simulated 4 different data sets spanning a range of dwell times from 10 seconds to 5 minutes.  By dwell time, we mean the notional time that an observer has spent collecting data in each cell, using an instrument like the one used to collect the data on the University of Texas campus. As with the real data, we binned all observations into 2048 equally spaced energy channels.  All methods operated on the counts within the bins, rather than with the raw simulated data.

\subsection{Results}

\begin{figure}[t]
\centering
\begin{subfigure}[t]{.47\textwidth}
\includegraphics[width=\textwidth]{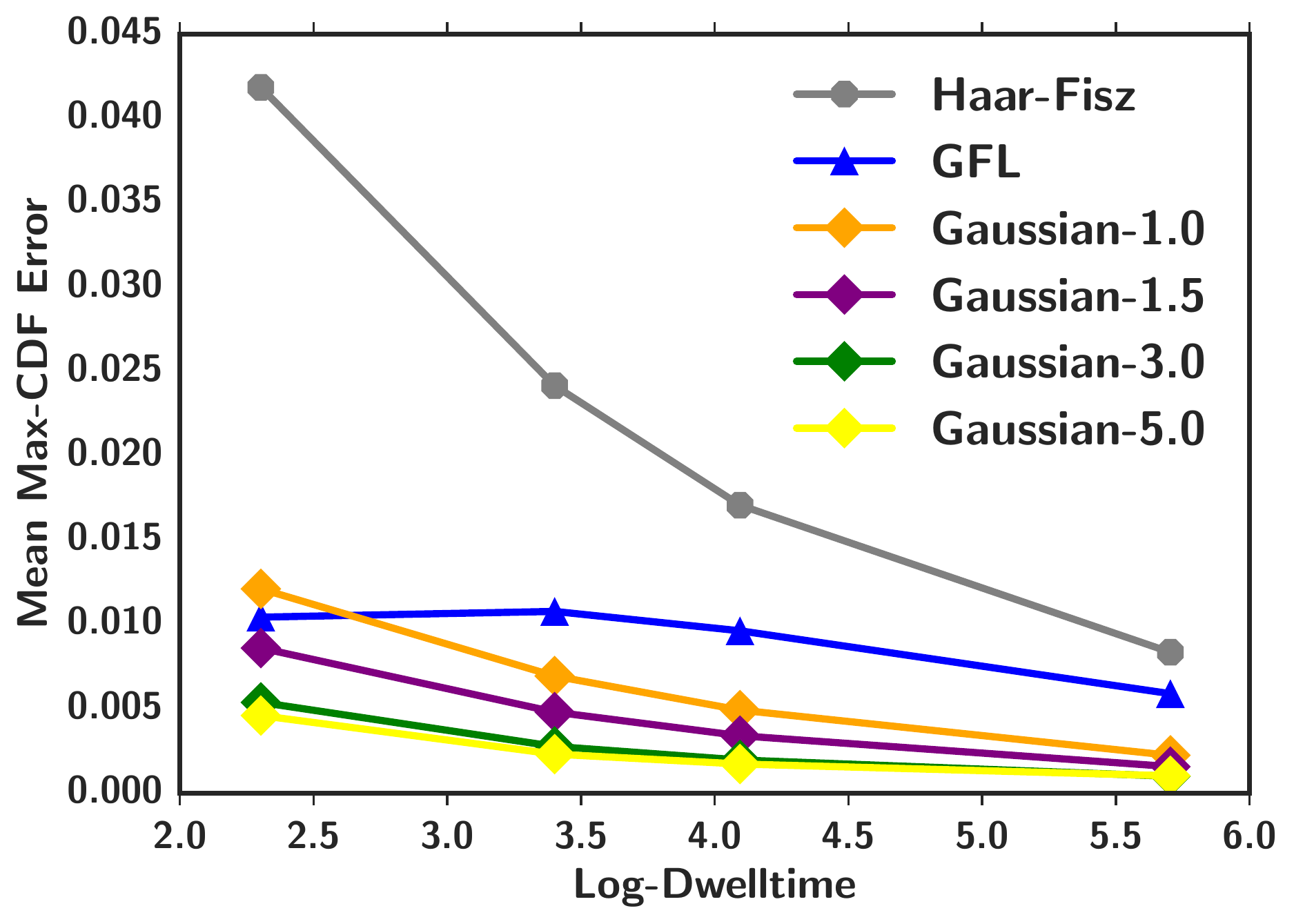}
\end{subfigure}\qquad
\begin{subfigure}[t]{.46\textwidth}
\includegraphics[width=\textwidth]{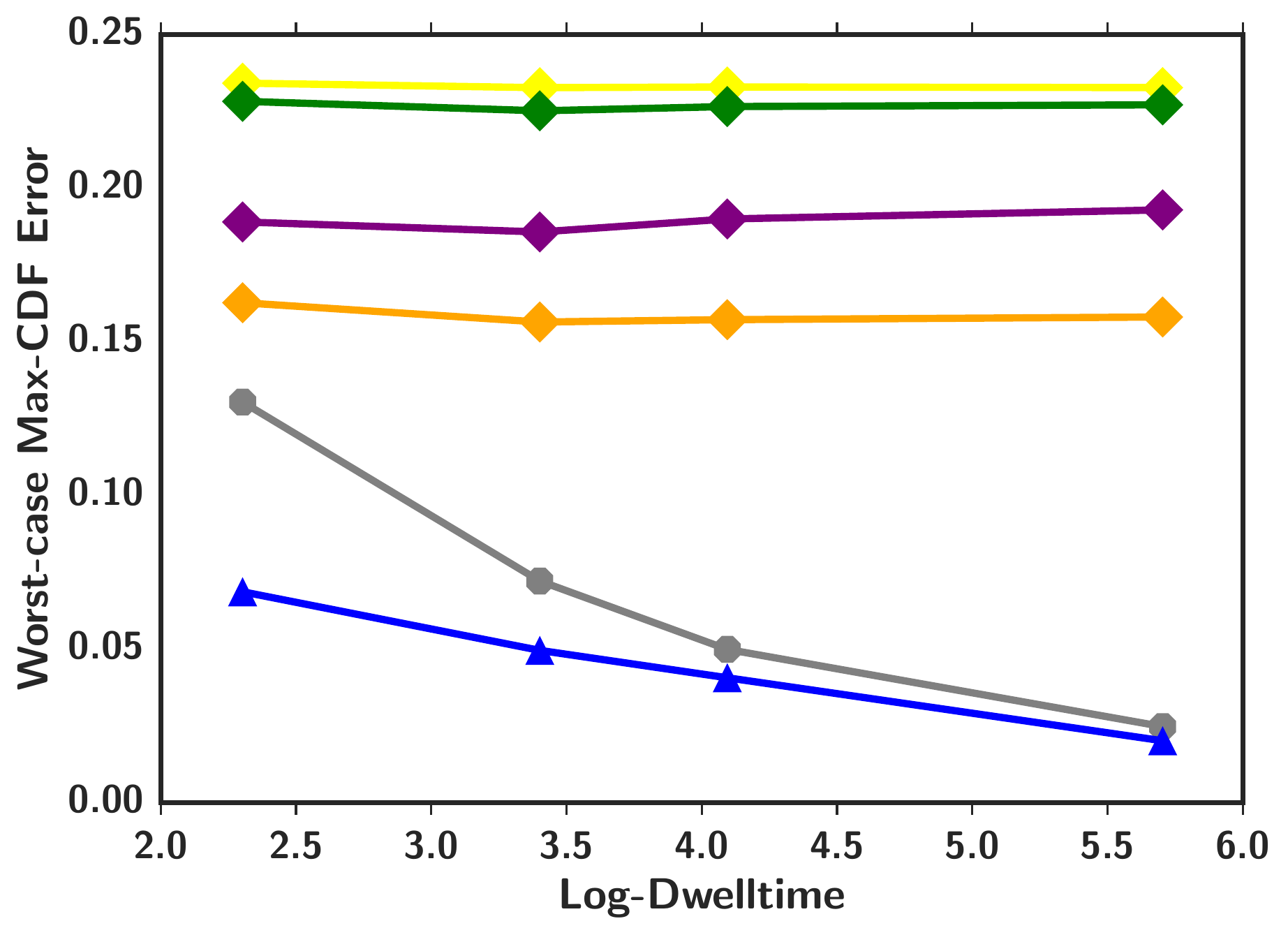}
\end{subfigure}
\caption{\label{fig:sim_study_results} Results of the large-scale spatial simulation.  The left panel shows average error across all spatial cells, while the right panel shows worst-case error.  Gaussian kernel smoothing with a large bandwidth leads to superior average performance, given the nearly stationary nature of the problem.  But it does so at the cost of drastically higher worst-case performance, regardless of how much data there is.}
\end{figure}

\begin{figure}[t]
\centering
\parbox[b]{0.04\textwidth}{\rotatebox{90}{\hspace{0.2in}(100, 100)}}
\begin{subfigure}[t]{.23\textwidth}
\includegraphics[width=\textwidth]{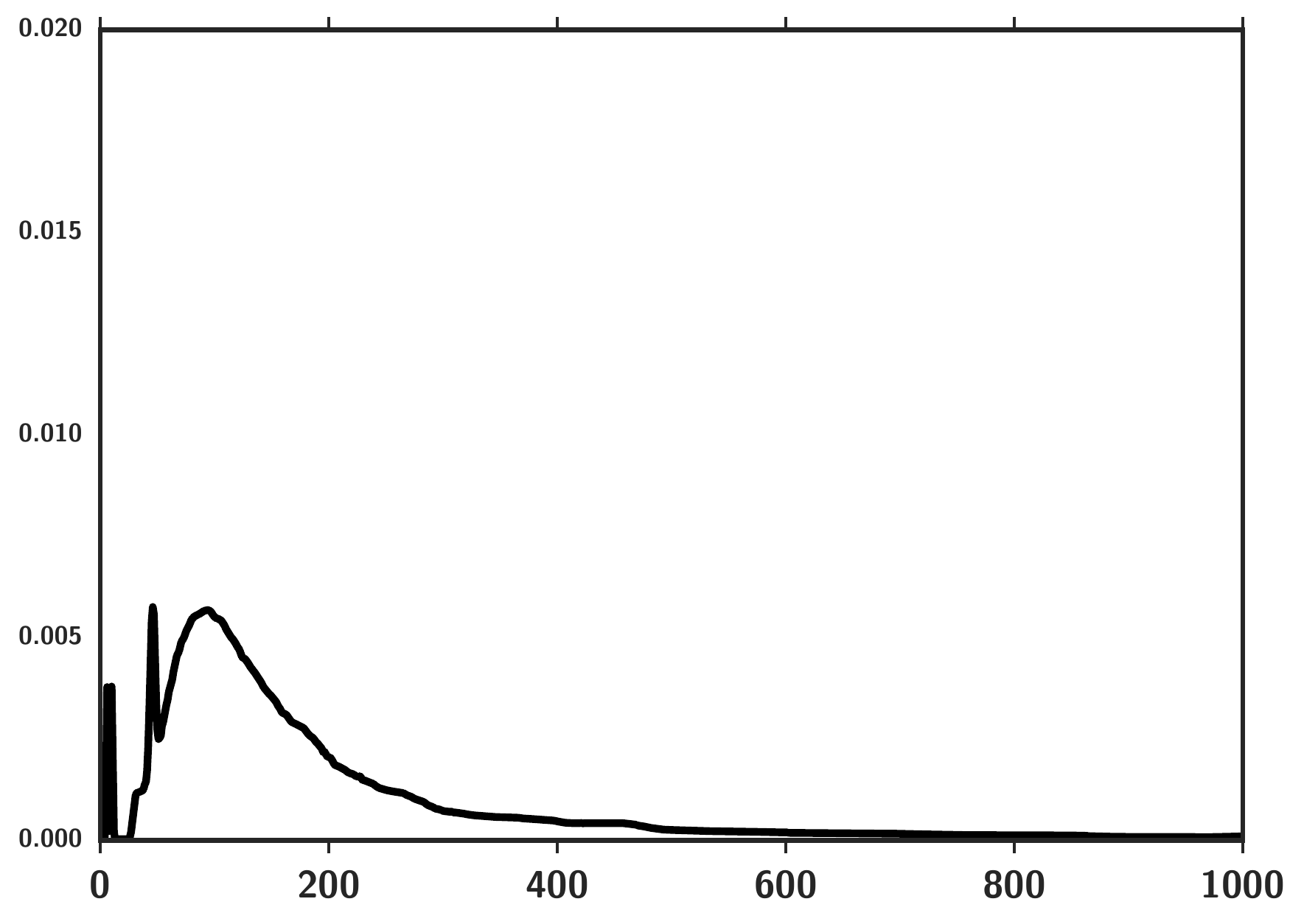}
\end{subfigure}
\begin{subfigure}[t]{.23\textwidth}
\includegraphics[width=\textwidth]{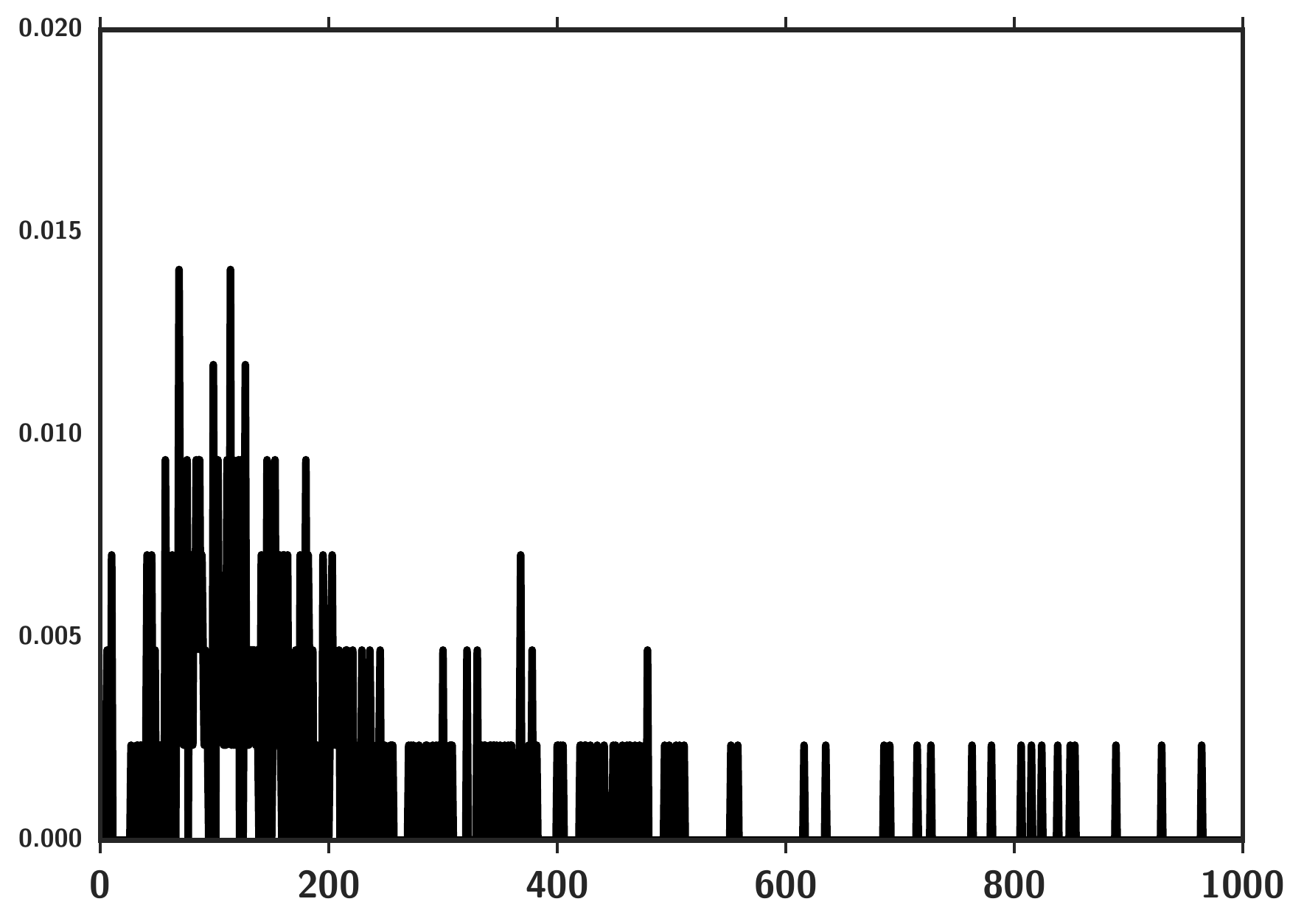}
\end{subfigure}
\begin{subfigure}[t]{.23\textwidth}
\includegraphics[width=\textwidth]{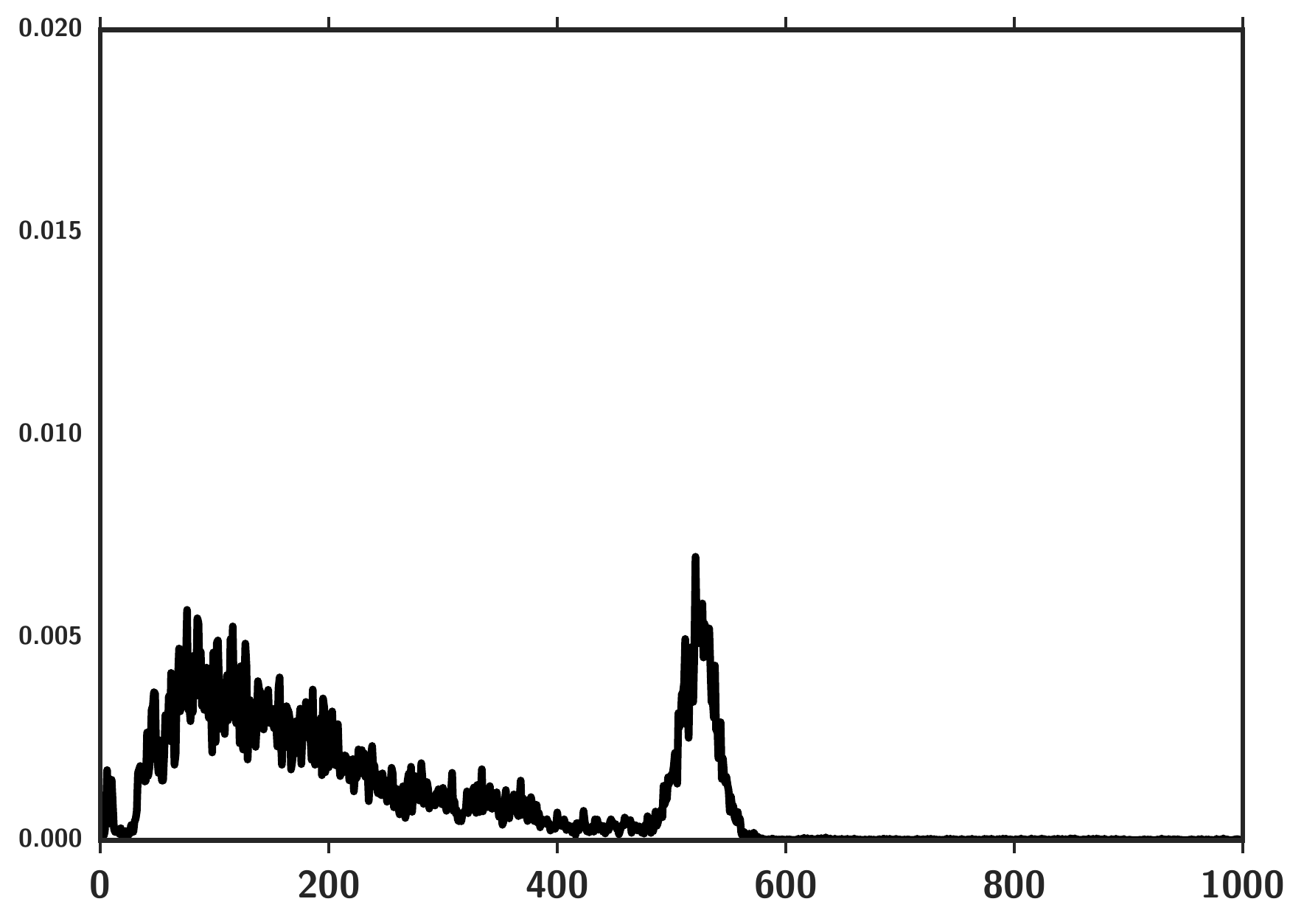}
\end{subfigure}
\begin{subfigure}[t]{.23\textwidth}
\includegraphics[width=\textwidth]{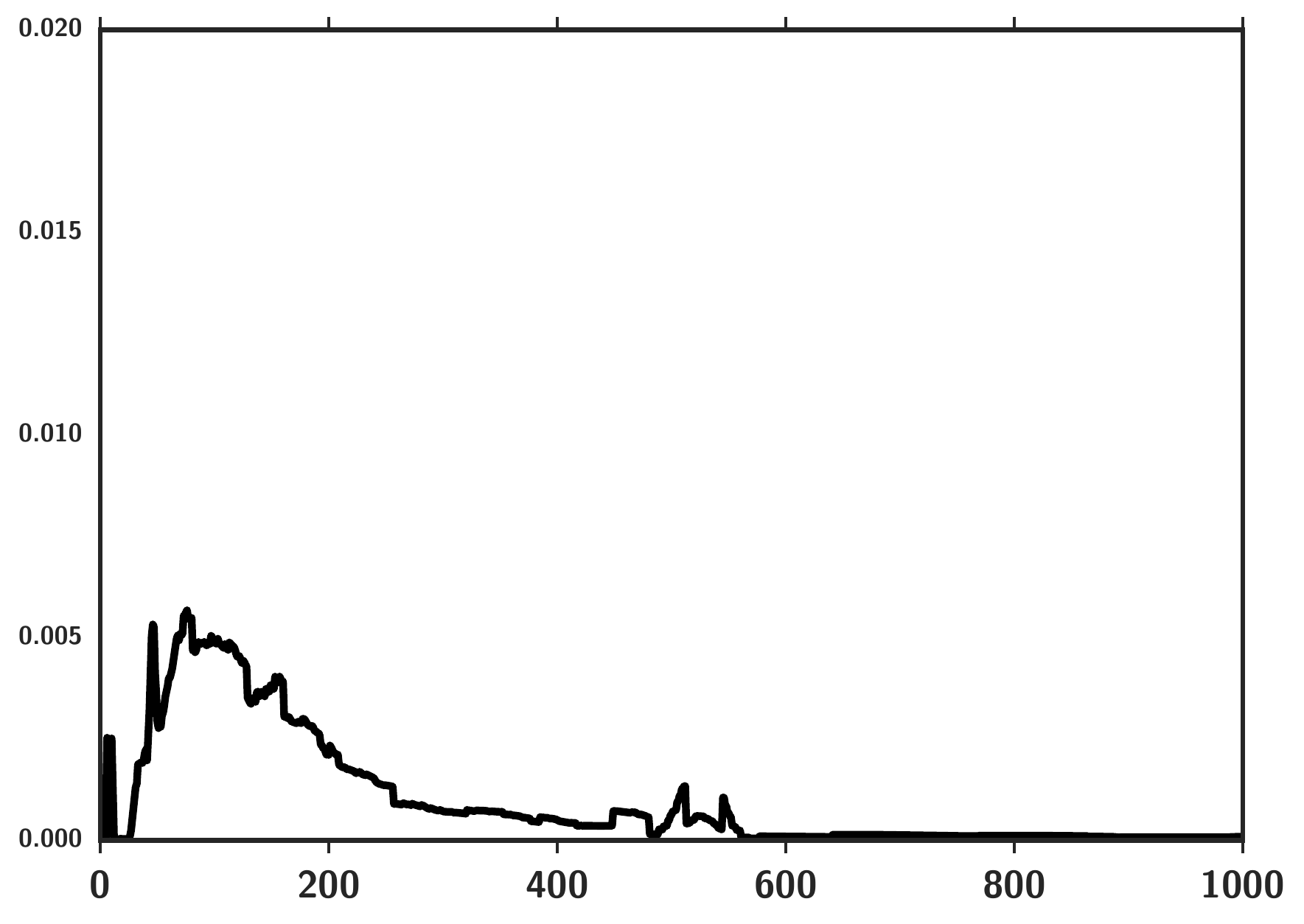}
\end{subfigure}
\parbox[b]{0.04\textwidth}{\rotatebox{90}{\hspace{0.2in}(101, 101)}}
\begin{subfigure}[t]{.23\textwidth}
\includegraphics[width=\textwidth]{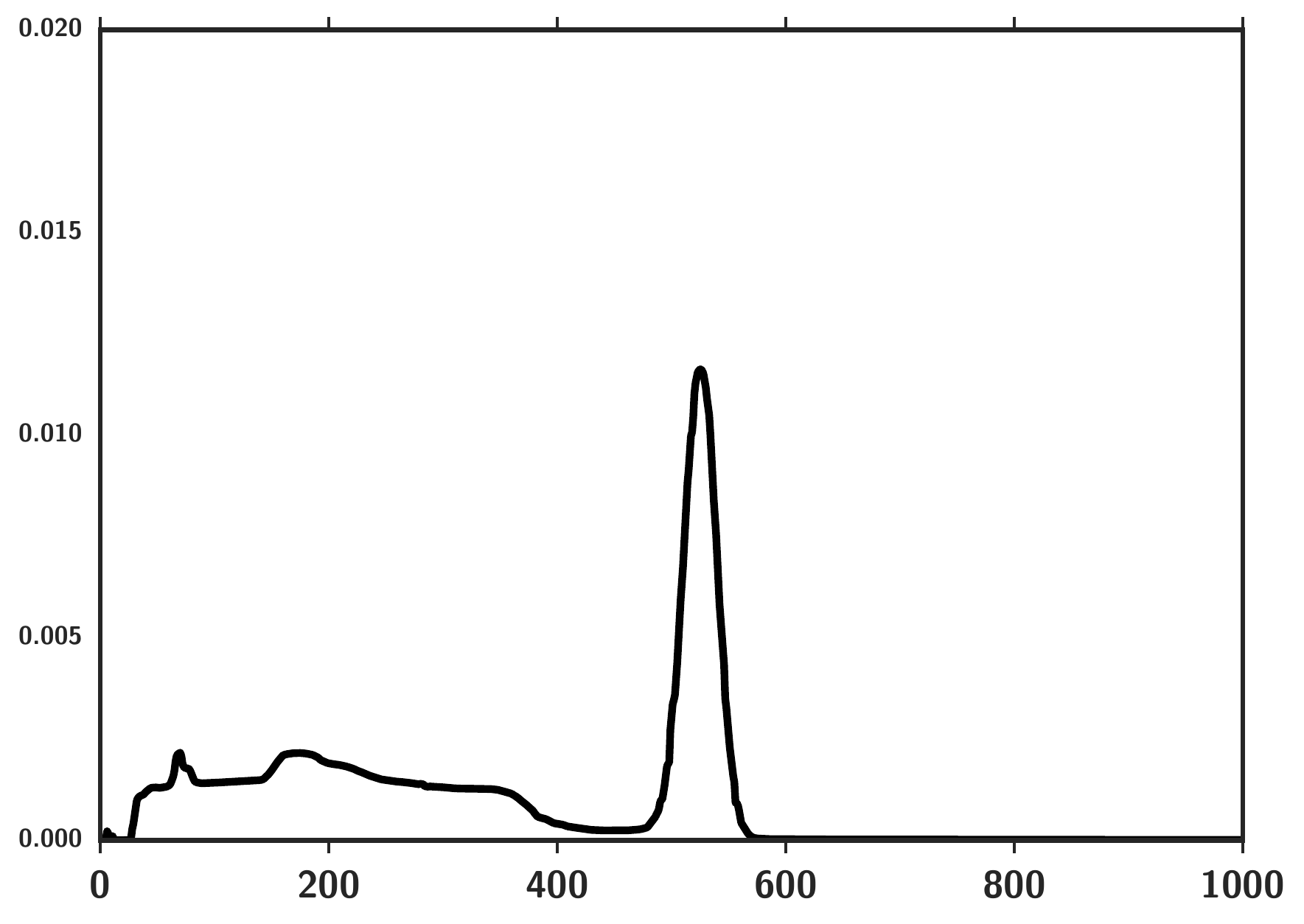}
\caption{Truth}
\end{subfigure}
\begin{subfigure}[t]{.23\textwidth}
\includegraphics[width=\textwidth]{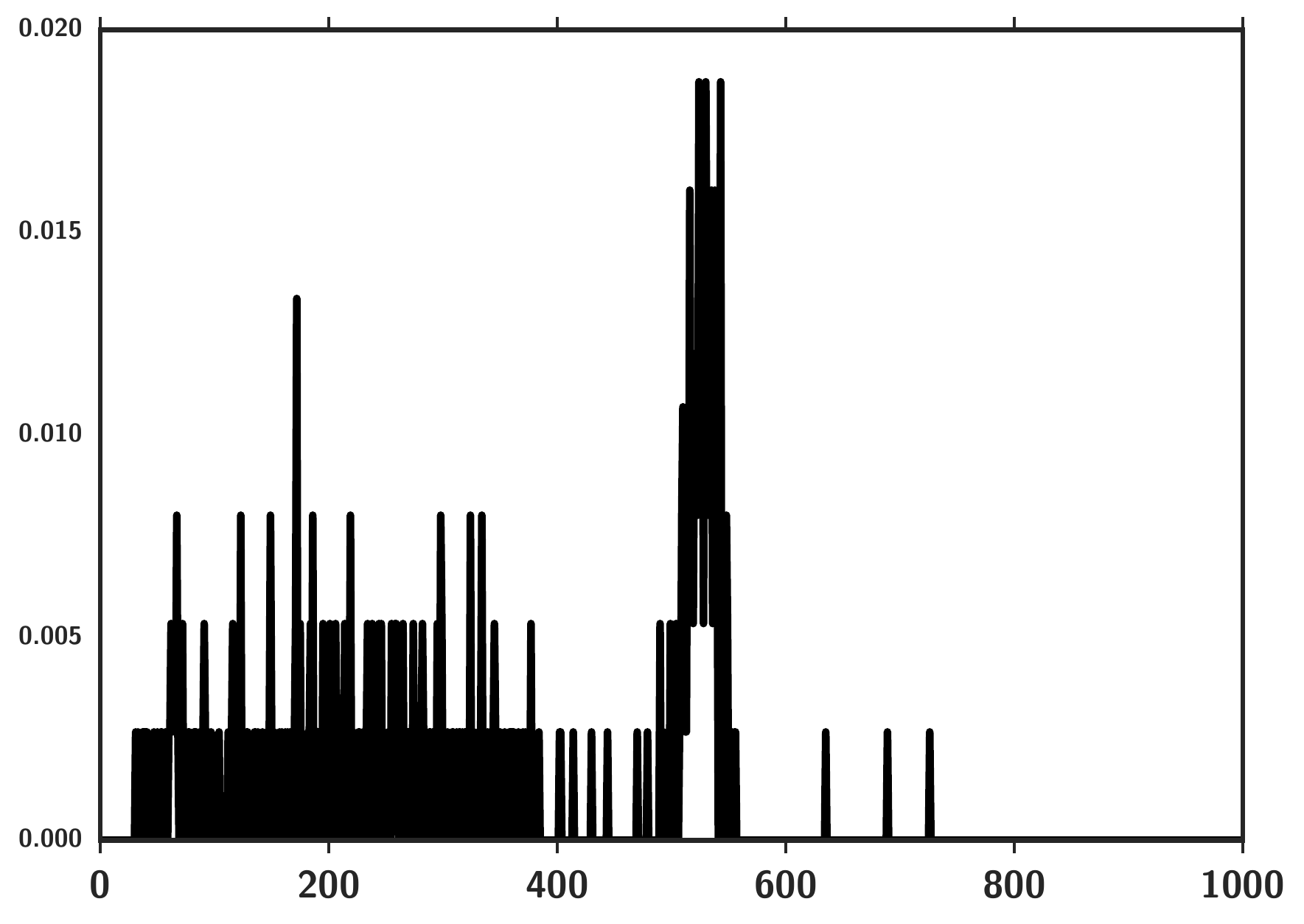}
\caption{Haar-Fisz}
\end{subfigure}
\begin{subfigure}[t]{.23\textwidth}
\includegraphics[width=\textwidth]{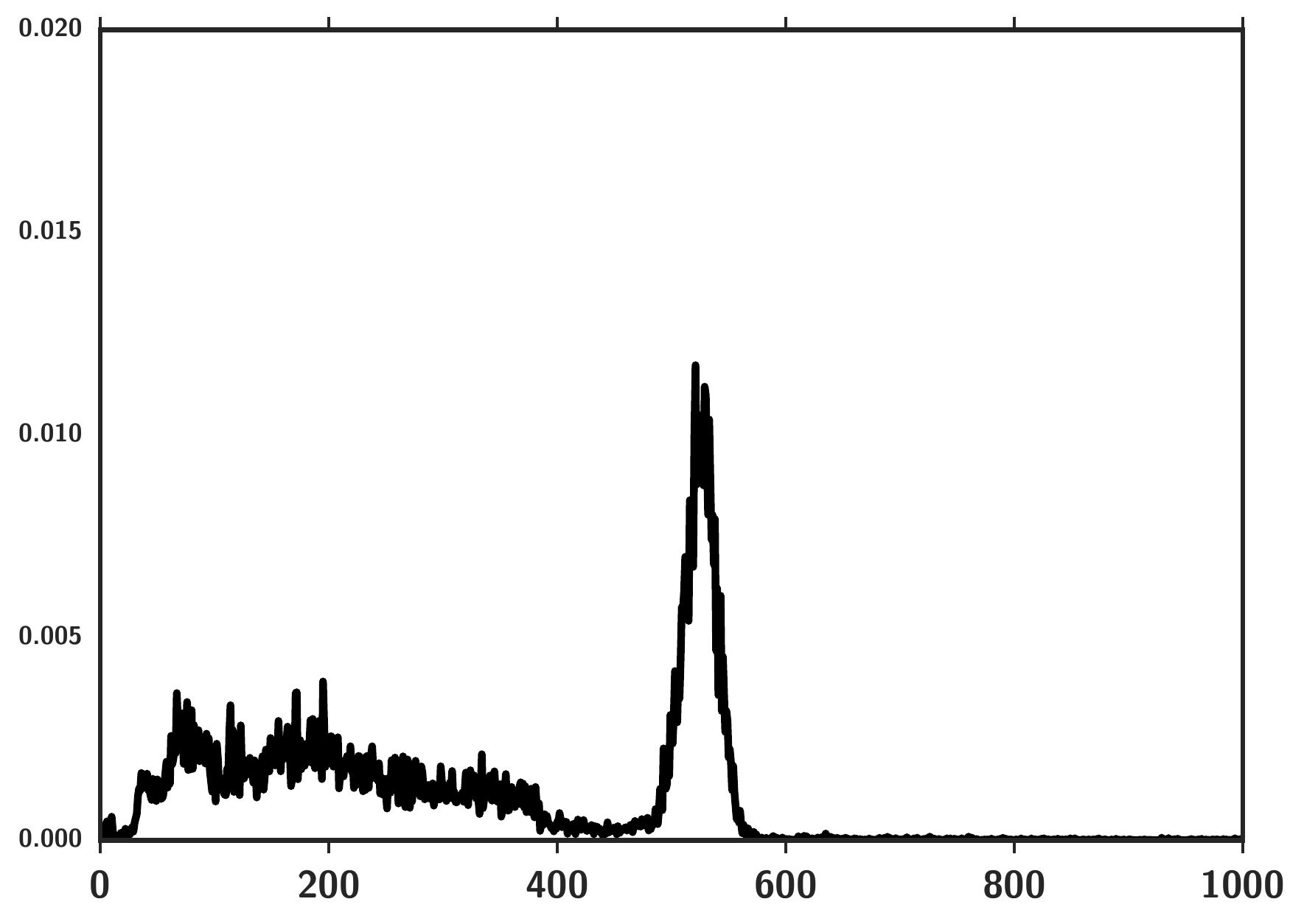}
\caption{Gaussian (c=5)}
\end{subfigure}
\begin{subfigure}[t]{.23\textwidth}
\includegraphics[width=\textwidth]{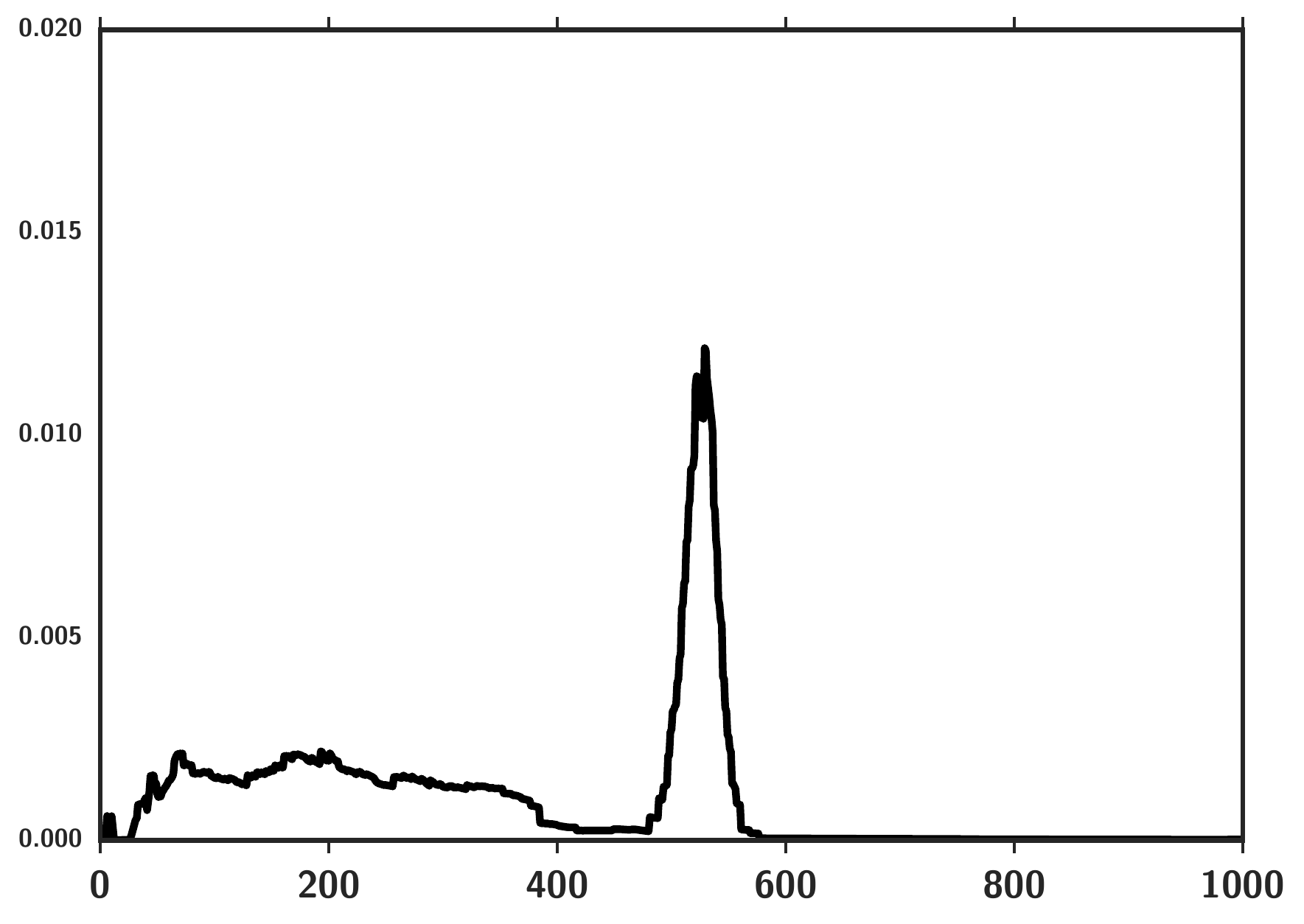}
\caption{GFL}
\end{subfigure}
\caption{\label{fig:sim_study_reconstructions} Example reconstructed densities on opposite sides of the occlusion, based on 10 seconds of data.  Cell (100,100) on the top row, while cell (101,101) is on the bottom row.   The Haar-Fisz model ends up doing no smoothing at all in these cells, because its adaptively chosen threshold parameter retained all wavelet coefficients in the Haar-Fisz expansion of the underlying density.  The Gaussian model is not able to adapt to the discontinuity, resulting in a spurious spike in the top density. In contrast, the GFL method is able to borrow statistical strength from neighboring cells, yielding a much smoother estimate, while simultaneously detecting the sharp discontinuity caused by the occlusion. }
\end{figure}

Using this simulated data, we benchmarked multiscale spatial density smoothing against two alternatives.
\begin{description}
\item[Haar-Fisz smoothing.] We applied the wavelet-based method of \citet{fryzlewicz:nason:2004} to estimate the density independently within each spatial cell.  This involves no spatial smoothing, but does denoise the observations within each cell in a way that attempts to preserve both spiky and smooth features.  The threshold parameter of the Haar--Fisz method was chosen by the greedy tree-based algorithm implemented in the \verb|haarfisz| R package \citep{fryzlewicz:2009}.  See \citet{jansen:2006} for further discussion and details of the method.
\item[Multiscale Gaussian kernel smoothing.] We also applied the standard technique of Gaussian kernel smoothing to yield a spatially smooth set of estimates for the splitting probabilities (Equation \ref{eqn:binomial_likelihood}) in each node in the multiscale tree decomposition.  Gaussian kernel smoothing is in widespread use for scalable spatial smoothing in a wide variety of fields, from fMRI to climate modeling.  In our case, the estimate at each node is replaced with a locally-weighted average of its neighborhood, with each neighbor's weight proportional to $\exp(-\frac{1}{2}\delta(s, t) / c) \cdot n_t \, ,$ where $s$ and $t$ are the node locations corresponding to the target and neighbor nodes, respectively, $\delta$ is the Euclidean distance function, $n_t$ is the number of trials observed at node $t$, and $c$ is a bandwidth. We explored bandwidth settings of 1, 1.5, 3, and 5 cells, truncating the kernel weights to zero after $3c$ cells away from the target node to yield a sparse smoothing matrix.  
\end{description}
Unfortunately, on a problem this large (250,000 spatial sites), the Bayesian approaches described earlier cannot feasibly be applied as benchmarks: they take hours to run on a problem with only 2,500 nodes, and their computational requirements scale much faster than linearly as the number of nodes increases.

Figure \ref{fig:sim_study_results} shows the average and worst-case error across all sites for all the methods, as measured in the previous experiment.  As expected, because the truth is nearly stationary, the Gaussian kernel-smoothing estimates have the lowest average error; higher bandwidths lead to more smoothing, and perform better on average (left panel).  However, the right panel of Figure \ref{fig:sim_study_results} shows that there is an high price to be paid for this superior average performance.  The worst-case error (i.e.~the worst estimate across all spatial sites) of the Gaussian kernel smoother stays nearly constant even as data accumulates, regardless of kernel bandwidth.  Moreover, for the Gaussian methods, there is an inverse relationship between average performance and worst-case performance.  For example, a bandwidth of 5 leads to the best average performance across all cells.  But even with 5 minutes of data, its worst reconstruction has an error of nearly 0.25, which is very poor given that the maximum possible CDF error is 1.   The worst-case error of the GFL-based method, on the other hand, gets smaller with more data.

REturning again to Figure \ref{fig:sim_study_corner}, we note that this poor worst-case performance of the Gaussian method is not isolated to one or two cells.  Panels B-D show the reconstruction error for each method for all cells from (71,71) to (141, 141), in the region where there is discontinuity.  Both Haar-Fisz smoothing and the graph fused lasso method have errors that are nearly spatially invariant in this region.  The Gaussian kernel smoother, on the other hand, has very poor performance along the entire ridge of the discontinuity.  This worst-case performance does not get any better as data accumulates.

Figure \ref{fig:sim_study_reconstructions} further highlights the effects of the discontinuity by zooming in on the reconstructions at two nearby cells on opposite sides of the occlusion, for a dwell time of 10s. The Haar-Fisz model does no smoothing at all here, because its adaptively chosen threshold parameter retained all wavelet coefficients in the Haar-Fisz expansion of the underlying density. The Gaussian kernel smoother estimates a false spike in the middle of the density at (100,100), because it is unable to adapt to the occlusion. The GFL method, on the other hand, reconstructs both densities with high fidelity.

On the basis of these experiments, we conclude that the GFL method is clearly the best choice for spatial smoothing in large-scale radiological survey.  Even in a (highly unfavorable) near-stationary setting, its average reconstruction is comparable to Gaussian kernel smoothing, albeit slightly worse.  But its worse-case reconstruction error is much better, because of the discontuity.  This makes the GFL method far more suitable for use in our application to spatially aware radiological anomaly detection described in the next section.  Global linear shrinkage is clearly inappropriate for this problem: its worst-case errors cannot be remedied with more data (except by not smoothing at all), and it therefore creates a situation where false alarms are practically guaranteed in any spatial region with a sharp change in the background.

\section{Radiological survey and anomaly detection at UT-Austin}
\label{sec:anomaly_example}

\label{sec:main_application}

\subsection{Protocol}

As discussed in the introduction, the main motivation behind our multiscale spatial density smoother is to allow improved detection of radiological anomalies.  We now show the proposed method achieves this goal, using the 18 hours of data from the Pickle Research Campus described in the introduction.

True radiological anomalies are rare. This is good from the standpoint of public health and safety.  But it does present a serious challenge when designing a study to assess the effectiveness of a method for radiological anomaly detection.  Without the budget or personnel for a large field study involving actual radioactive materials and complicated observational protocols, we are forced to rely on simulated anomalies.  Our focus, therefore, is on making them as realistic as possible, by leveraging two key sources of data: (1) the 18 hours of normal background data for the UT Pickle Research Campus, as described in the introduction; and (2) the experimental data collected on the empirical spectra of cesium and cobalt sources, as described in the Appendix.

\paragraph{Training.}  We first split the campus background data into training ($\approx 80\%$) and testing ($\approx 20\%$) sets.  Specifically, the training data had 2,015,515 total photon counts, and the testing data 546,064 total photon counts, across all energy channels.  We partitioned the campus into a regular grid of 50m $\times$ 50m spatial sites.  Using only the training data, we constructed site-specific estimates of the gamma-ray spectrum via multiscale density smoothing, as described in Sections \ref{sec:preliminaries} and \ref{sec:spatial_smoothing}.  We also used the training data to estimate a global (spatially invariant) background density, simply by averaging the counts in each bin across all training observations.  We refer to these as the local and global estimates, respectively.  The data were far from uniformly distributed across the spatial sites: in the training data, the average site had 12,140 photon counts, the data-richest site had 249,000 counts, while the data-poorest site had only 64 counts.

Figure \ref{fig:prc_reconstruction_examples} shows two example estimates for sites that are separated by several hundred meters, together with the global estimate and the raw data at those sites.  The training data for these two sites exhibit visibly different backgrounds near the broad peak in the spectrum between channels 50 and 200; compare each histogram with the left two panels using the dashed black line (the global average background) as a common point of reference.  Our procedure faithfully reconstructs these differences without the excessive variability of the raw empirical spectra.

\paragraph{Testing.}  Suppose we are trying to simulate $T$ seconds of observation from site $s$.  We assume that, in addition to the normal background radiation, the gamma-ray spectrum also includes the emissions from an anomalous source with gamma-ray spectrum $f_a$, of size $m$ milliCuries, and at distance $d$ meters from the observer.  To generate such an observation, we used the testing data as follows.  A key feature of this experimental design is that we never assume a known background.
\begin{enumerate}[(1)]
\item To simulate the background contribution, sample $T$ one-second background observations $x^{(s)}_{[1]}, \ldots, x^{(s)}_{[T]}$ by bootstrapping the subset of the testing data that came from site $s$ (i.e.~sampling its rows with replacement).  Aggregate these observations to form a single simulated $T$-second contribution $b^{(s)}$ from the background:
$$
b^{(s)} = \sum_{t=1}^{T} x^{(s)}_{[t]} \, .
$$
The summation is component-wise: both the bootstrapped observations $x^{(s)}_{[t]}$ and the summand are vectors of length 2048.

\item To simulate the contribution of the anomaly, use Equation (\ref{eqn:sizetoCPS}) to convert the source size $s$ and distance $d$ to an expected per-second count rate $\lambda$.  Sample a Poisson random variable $n \sim \mathrm{Poisson}(T \lambda)$ for T seconds of observation.  Then sample $n$ photons from the known source density $f_a$ (e.g.~the right two panels of Figure \ref{fig:bg_co_cs_densities}).  Aggregate these samples into the 2048 energy channels to form the contribution from the anomalous source, $a^{(s)}$.

\item Combine the contributions from background and anomaly,
$$
\tilde{x}^{(s)} = b^{(s)} + a^{(s)} \, ,
$$
to form a single ``bootstrap + anomaly'' count vector that simulates $T$ seconds of observation at site $s$.

\end{enumerate}

The goal of our study is to see how easily our method can distinguish such an $\tilde x^{(s)}$ from normal (background-only) observations by comparing it to the background spectrum estimated from the training data.

\paragraph{Benchmarks and simulation settings.}

\begin{figure} \begin{center}
    \includegraphics[width=6in]{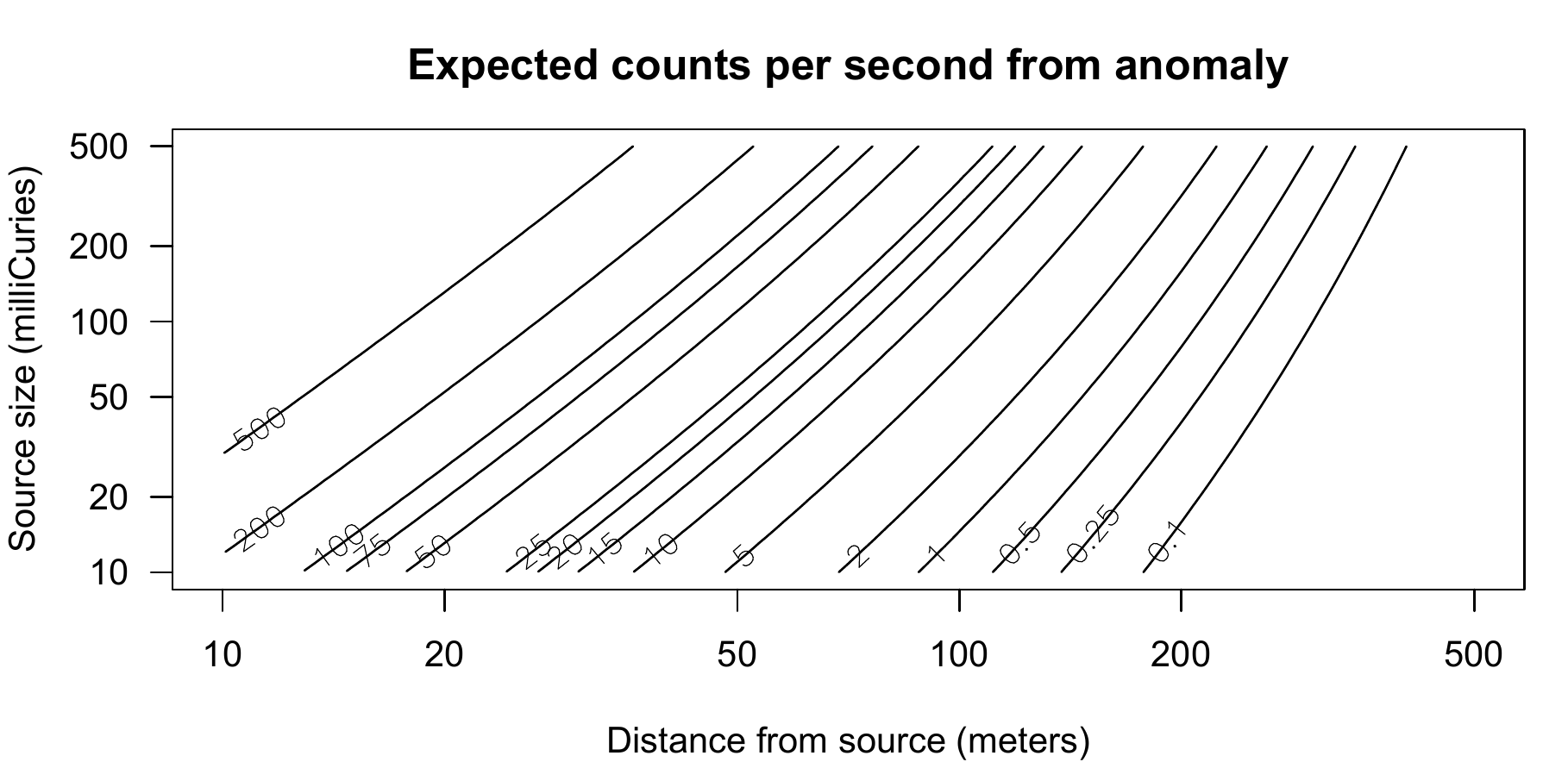} \\
    \bigskip \includegraphics[width=6in]{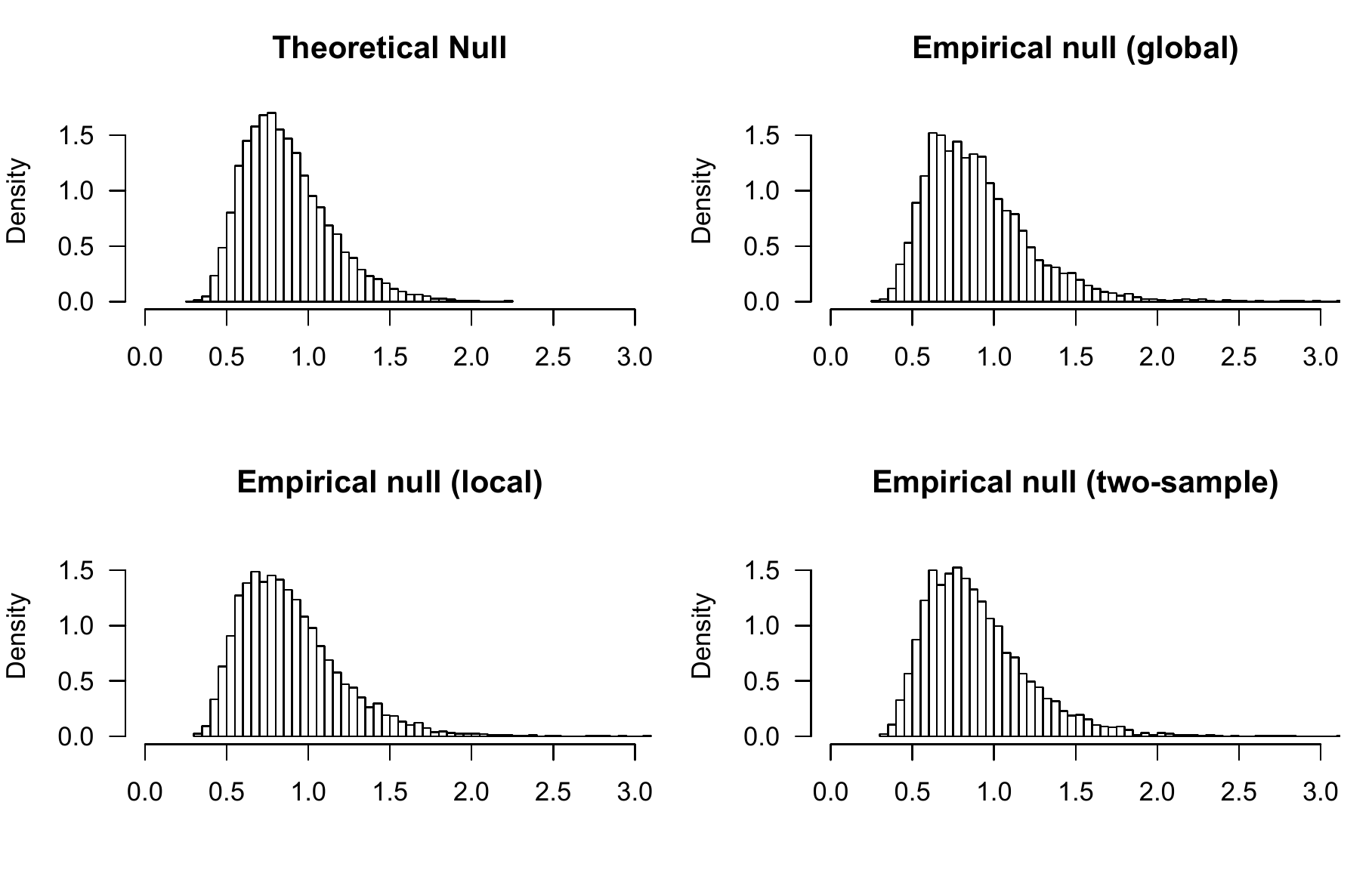} \caption{\label{fig:null_comparison_and_cps} Top panel: the relationship between source size, distance to observer, and expected counts per second (shown as labeled contours) from an anomalous source.  Expected counts per second refer to expected counts measured by the detector used in our study, not to any measure of intrinsic strength of signal.  Bottom four panels: comparison of the theoretical KS null distribution with the three empirical null distributions estimated from the testing data and used to compute the false positive rate in our estimated ROC curves.}
\end{center}
\end{figure}

The goal of this paper is not to invent a new anomaly-detection algorithm, but rather to show that spatial smoothing can improve existing algorithms.  Therefore, to decide whether a simulated $\tilde x^{(s)}$ is an anomaly, we use a one-sample Kolmogorov--Smirnov test comparing the empirical energy CDF of $\tilde x^{(s)}$ to the CDF of the estimated background spectrum at site $s$.  This is identical to the method of \citet{chan:etal:2014} with overdispersion parameter $\phi=1$; the generalization to other overdispersion parameters would be straightforward in principle, although we do not explore this.\footnote{Note that the authors of \citet{chan:etal:2014} simply assume that the background is known.}

To establish that spatial disaggregation and smoothing can improve matters, we benchmarked our method against two others:
\begin{enumerate}[(1)]
\item The one-sample KS/\citet{chan:etal:2014} test using the global estimate of the background as the reference distribution.
\item The two-sample KS test of \citet{Reinhart:2015}, where ``sample 1'' comprises the training observations from site $s$, and ``sample 2'' is the simulated $\tilde x_{(s)}$.
\end{enumerate}

The simulation parameters to be varied are the source size $m$, the source distance $d$, the source spectrum $f_a$, and the observation time $T$.  However, the mapping from source size and distance to expected count rate (Equation \ref{eqn:sizetoCPS}) is many-to-one.   (The top panel of Figure \ref{fig:null_comparison_and_cps} shows this relationship.) We therefore parametrize our results in terms of the expected anomaly count rate rather than size and distance individually.  We simulated data across all combinations of the following settings:
\begin{itemize}
\item time $T \in \{2,  4,  6,  10,  14,  20,  30, 40,  60,  90, 120, 180\}$ seconds.
\item anomaly count rate $r \in \{0.1, 0.25, 0.5, 1, 2, 5, 10, 15, 20, 25, 50\}$ photons per second.
\item $f_a \in \{f_{\mathrm{Cs}}, f_{\mathrm{Co}} \}$.  See Figure \ref{fig:bg_co_cs_densities}.
\end{itemize}

For each combination of settings, we used the bootstrap-based procedure just described to simulate 1000 anomalous observations $\tilde x^{(s)}$ at each site.  For a range of detection thresholds $t$, we computed the true positive rate of each procedure, i.e.~the fraction of simulated $\tilde x^{(s)}$'s whose KS statistics exceeded $t$.  We also computed the false positive rate, i.e.~the probability that the KS statistic for a normal observation will exceed $t$.  For each method, we aggregated the false positives and true positives across all spatial sites to compute a method-specific ROC curve for every simulation setting.

\paragraph{Choice of null hypothesis.}  To be conservative, we did not compute the false positive rate for each method using theoretical reference distributions for the one- or two-sample KS statistics.  Rather, we calculated an ``empirical null'' for each method by applying each KS test to the testing data without artificial anomalies injected.  These empirical null distributions, together with the standard KS reference distribution, are shown in the bottom four panels of Figure \ref{fig:null_comparison_and_cps}.  The empirical nulls are close to the theoretical null but have slightly thicker right tails, since the one-sample KS test does not account for uncertainty in the background estimates.

\subsection{Results}

\begin{figure}[t]
\begin{center}
\includegraphics[width=\textwidth]{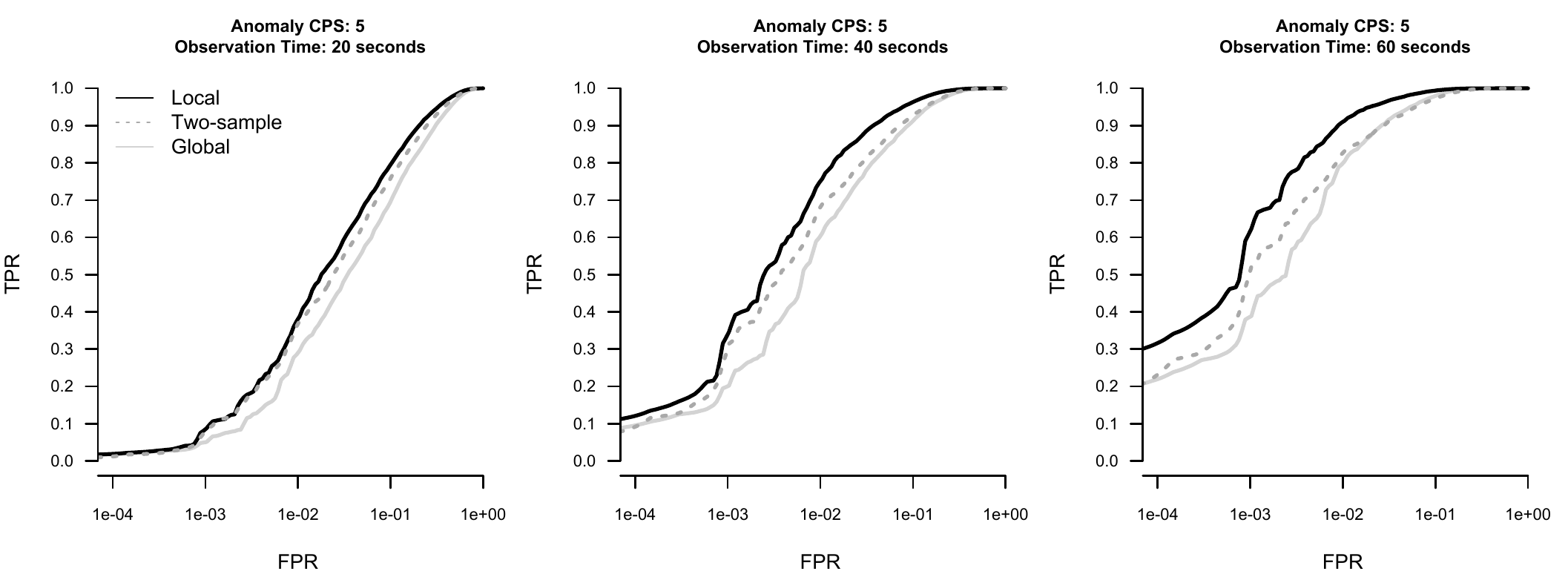}
\caption{\label{fig:prc_anomaly_ROC} ROC curves for detecting a 5 photon/second cesium source under varying observation times.  The black line shows the ROC curve when using site-specific spectral densities estimating via multiscale spatial density smoothing.  The solid grey lines show the ROC curve when using a global (spatially invariant) estimate of the spectral density.  The dotted grey lines show the ROC curves using the two-sample KS test described in \citet{Reinhart:2015}, which also adapts spatially but does not involve any spatial smoothing.  In all panels, the horizontal axis (FPR, false positive rate) is shown on a log scale, while the vertical axis (TPR, true positive rate) is shown on an ordinary scale.}
\end{center}
\end{figure}

Across all simulation settings, the KS test based on spatially smoothed density estimates yielded uniformly better ROC curves than the other methods.  The one-sample KS test based on a global density estimate performs the worst, while the two-sample KS test is intermediate between the two.  This demonstrates nicely that there are two distinct sources of advantage here: one performance bump due to spatial disaggegration, and another distinct bump due to spatial smoothing.

Figure \ref{fig:prc_anomaly_ROC} provides a small sample of our results.  In these three panels, the expected count rate from the anomalous source is 5 gamma rays per second, compared to a background rate of 39 per second, and the observation time increases from 20 to 40 to 60 seconds reading left to right.  A careful comparison of the panels shows that the KS test with local (spatially smoothed) density estimates performs essentially as well with 40 seconds of data as the two-sample method does with 60 seconds of data, suggesting a roughly 30\% improvement in time-to-detection.  Similar improvements were apparent across all simulation settings.

\section{Conclusions}
\label{sec:conclusions}

We have presented multiscale spatial density smoothing as a new technique for estimating a spatially varying density function.  Our evidence suggests that the method offers state-of-the-art performance when incorporated into anomaly-detection protocols for radiological survey data.  The power improvements we found over existing methods are no small matter in practical terms, especially in a law-enforcement context (where detecting a source with high-confidence 30\% faster may make a big difference).  Moreover, the simulation evidence of Section \ref{sec:simulations} show clearly that, at least for radiological surveys, spatial smoothing using our method yields a favorable blend of good average error with good worst-care error.  It is also clear that global linear shrinkage (whether in the form of a CAR model or Gaussian kernel smoothing) cannot handle the kind of sharp changes in the background that occur in real radiological data.

One potential shortcoming is that, while our method smooths the density in space very effectively, it does not smooth the density across its underlying support $B$ in an especially sophisticated way.  (As it stands, the only way to incorporate this form of smoothing is to use a shallower tree composition, which has an effect similar to that of using a wider bin width in a histogram.)  Thus an obvious way to generalize the method would be to merge our work with some of the ideas discussed by, for example, \citet{jansen:2006} and \citet{willett:nowak:2007}, to smooth along the energy spectrum.  (None of these techniques incorporate spatial smoothing.)  It is not immediately obvious to us how to accomplish this synthesis, but this is an active area of research.  

\paragraph{Acknowledgements.}  The authors thank Patrick Vetter of the UT Applied Research Laboratories for his assistance with the pilot studies described here; the University of Texas Police Department for their ongoing collaboration with data collection; and Ryan Tibshirani of CMU for sharing his expertise on algorithms for the graph-fused lasso.

\appendix

\section{Pre-processing}
\label{app:remarks}

In this appendix, we provide on further details about our data and pre-processing.  There are some notable features of our data that almost certainly do not correspond to true physical features of the underlying gamma-ray spectrum.  Rather, they correspond to quirks of the equipment being used as part of the radiological survey.  These ``detector artifacts'' are distinct from the true anomalies that we hope to be able to find (i.e.~departures from the background due to the presence of a radioactive source).  

Because most background and man-made radiation occurs at lower energies, we used only the lower 2,048 bins of the recorded spectra, Winsorizing the data so that all gamma rays at higher energies are placed in the 2,048th bin. This causes a noticeable peak in this bin. There are also two very low-energy peaks (between bins 5 and 10) which do not appear to be natural, and may be an artifact of our detector. Nonetheless, these artifacts are ``real'' from the standpoint of a statistician modeling the data, and can be expected in any real-world data.

To avoid these artifacts when detecting anomalies, we use only the data from channels 16 to 2047. However, for the sole purpose of benchmarking our statistical models, we have decided to analyze the entire recorded spectrum. We make this choice partially to reduce the researcher degrees of freedom that inevitably arise in methodological benchmarks.  But we also do so to illustrate a major strength of our approach: it does not require close supervision by radiation physicists or spectrometer engineers. For a large-scale radiological survey, where cheap detectors and inexpert operators are likely to be the norm rather than the exception, such robustness emerges as an asset of our methodology.

\section{Protocol for estimating empirical spectra}
\label{app:protocol}

To obtain accurate cesium-137 and cobalt-60 spectra, we used small check sources provided by the University of Texas's Nuclear Engineering Teaching Laboratory. The detector was partially shielded from the natural background radiation using lead bricks and the check source placed 5 centimeters away. The detector then recorded several minutes of gamma rays, producing a detailed spectrum.

The cesium-137 source was used to calibrate Equation \eqref{eqn:sizetoCPS}. The check source contained 844 nanoCuries of cesium-137 and resulted in an average of 630 gamma rays per second at the detector, observed over several minutes. Gamma radiation is known to follow a $1/r^2$ falloff in space with an additional exponential decay term due to absorption in air; hence, as a function of distance $r$, observed counts follow the relationship
\[
c(r) \propto \frac{1}{r^2} \exp(-\mu r),
\]
where \(\mu\) is a known attenuation coefficient in air. For cesium-137, which emits gamma rays primarily at 660 kiloelectronvolts, we used \(\mu = 0.0100029\, \text{m}^{-1}\).

Given this proportionality and a known count rate at 5 centimeters, we can derive the ratio between the known count rate and the expected count rate at any distance. This produces Equation \eqref{eqn:sizetoCPS}. Because the attenuation coefficient varies with energy, it is an approximation.  Simulated count rates for cobalt-60, which emits at higher gamma ray energies, may be less accurate.

\section{Further details of Bayesian method}
\label{app:bayes}

The Bayesian version of binomial graph trend filtering, outlined in Section \ref{sec:bayesian_smoothing}, has the following structure.    Recall that $y^{(s)}$ is the count in the left-child node, and $m^{(s)}$ is the count in the parent node.  The goal is to estimate $w^{(s)}$, the conditional ``left split'' probability at the parent node.  To lighten the notation, we drop the subscript $\gamma$, indexing the tree node.
$$
\begin{aligned}
(y^{(s)} \mid m^{(s)}) &\sim \mbox{Binomial}(m^{(s)}, w^{(s)}) \, , \quad
w^{(s)} = \frac{ \exp\{ \alpha + \beta^{(s)} \} } {1 + \exp\{ \alpha + \beta^{(s)} \} } \\
p(\beta \mid \Omega) &\propto \exp \left\{ - \frac{1}{2} \bbeta^T (\Delta^T \Omega^{-1} \Delta) \bbeta \right\}  \\
\Omega^{-1} &= \text{diag}(\omega_1^{-1}, \ldots, \omega_D^{-1}) \\ 
(\omega_j \mid \nu_j) &\sim \mbox{Exp}(\nu_j^2/2) \\
\nu_j &\sim \mbox{Gamma}(1,\lambda) \\
p(\alpha) &\propto 1 \, .
\end{aligned}
$$
where $\lambda$ is the global shrinkage parameter.  Upon integrating out the latent variables $\omega_j$ and $\nu_j$, we find that each entry in the vector $\Delta \beta$ has a generalized double-Pareto or gamma-lasso prior \citep{armagan:etal:2013, taddy:2010}.

To handle the binomial likelihood, we introduce one Polya-Gamma latent variable $h^{(s)}$ for each node in the graph.  With this simple data-augmentation trick from \citep{polson:scott:windle:2012a}, the above model can be fit by Gibbs sampling, with the following conditionals.  We let $-$ denote all other variables being conditioned upon, and $\kappa^{(s)} = y^{(s)} - m^{(s)}/2$.  We also let $\mathbf{1}$ denote a vector of all 1's, and let $\delta_j = (\Delta \beta)_j$.
$$
\begin{aligned}
(h^{(s}) \mid -) &\sim \mbox{Polya-Gamma}(m^{(s)}, \psi^{(s)}) \, , \quad \psi^{(s)} = \alpha + \beta^{(s)} \\
(\beta \mid -) &\sim N(\mu_\beta, \Sigma_\beta)  \\
\Sigma_\beta^{-1} &= H + \frac{1}{2} \Delta^T \Omega^{-1} \Delta \\
\mu_{\beta} &= \Sigma_\beta^{-1}( \widetilde{\kappa} - \alpha \mathbf{1}) \, , \quad \widetilde{\kappa} = (\kappa_1/h_1, \ldots, \kappa_n/h_n) \\
\Omega^{-1} &= \text{diag}(\omega_1^{-1}, \ldots, \omega_d^{-1}) \\
(\omega_j^{-1} \mid -) &\sim \mbox{Inverse-Gaussian}(\sqrt{\nu_j^2/\delta_j^2}, \nu_j^2)\\
(\nu_j \mid -) &\sim \mbox{Gamma}(2,1 + |\delta_j|) \\
(\alpha \mid -) &\sim N(\mu_\alpha, \sigma^2_\alpha) \\
\sigma^2_\alpha &= \left( \sum h^{(s)} \right)^{-1} \\
\mu_\alpha &= \sigma^2_\alpha  \sum ( \kappa^{(s)} - \beta^{(s)}/h^{(s)}) 
\end{aligned}
$$
The multivariate normal draw for $\beta$ uses the technique of \citet{rue:2001}.  We center the $\beta$ vector after each draw, enforcing the constraint that $\mathbf{1}^T \beta = 0$.


\begin{small}
\singlespacing
\bibliographystyle{abbrvnat}
\bibliography{spatial_density_multiscale}

\end{small}

\end{document}